\documentclass[fleqn,usenatbib]{mnras}

\usepackage{newtxtext,newtxmath}

\usepackage[T1]{fontenc}
\usepackage{ae,aecompl}
\usepackage{bm}
\usepackage{color, colortbl}
\definecolor{Gray}{gray}{0.95}
\usepackage{multirow}
\usepackage{tabularx}

\usepackage{float}
\usepackage[caption = false]{subfig}
\usepackage{graphicx}	
\usepackage{amsmath}	
\usepackage{amssymb}	
\usepackage{booktabs}
\usepackage{psfig,epsfig}
\usepackage{natbib}
\usepackage{color}
\usepackage{multirow}
\usepackage{subfig}
\usepackage[title]{appendix}
\usepackage{float}
\usepackage{placeins}

\title[Cosmology with void abundances]
      {Cosmological exploitation of the size function of cosmic voids identified in the distribution
       of biased tracers}

\author[S. Contarini et al.] {Sofia Contarini$^{1,2}$\thanks{E-mail:
    sofia.contarini3@unibo.it}, Tommaso Ronconi$^{3,4}$, Federico Marulli$^{1,2,5}$, Lauro Moscardini$^{1,2,5}$, \newauthor Alfonso Veropalumbo$^{6}$ and Marco Baldi$^{1,2,5}$ \\ \\ 
    $^1$ Dipartimento di Fisica e Astronomia - Alma Mater Studiorum Universit\`{a} di Bologna, via Piero Gobetti 93/2, I-40129 Bologna, Italy\\ $^2$ INAF - Osservatorio di Astrofisica e Scienza dello Spazio di Bologna, via Piero Gobetti 93/3, I-40129 Bologna, Italy \\ $^3$ SISSA - International School for Advanced Studies, via Bonomea 265, I-34136 Trieste, Italy \\ $^4$ INFN - Sezione di Trieste, via Valerio 2, I-34127 Trieste, Italy \\ $^5$ INFN - Sezione di Bologna, viale Berti Pichat 6/2, I-40127 Bologna, Italy \\ $^6$ Dipartimento di Fisica, Universit\`a degli Studi Roma Tre, via della Vasca Navale 84, I-00146 Roma, Italy}

\newcommand{\plotone}[1]
           {\centering \leavevmode \psfig{file=#1,width=\columnwidth,clip=}}

\newcommand{\plottwo}[2]
           {\centering \leavevmode \psfig{file=#1,width=\columnwidth,clip=}
                            \hfill \psfig{file=#2,width=\columnwidth,clip=}}
\newcommand{\plotthree}[3]
           {\centering \leavevmode \psfig{file=#1,width=0.32\textwidth,clip=}
                            \hfill \psfig{file=#2,width=0.32\textwidth,clip=}
                            \hfill \psfig{file=#3,width=0.32\textwidth,clip=}}
\newcommand{\plotfull}[1]
           {\centering \leavevmode \psfig{file=#1,width=\textwidth,clip=}}

\begin{document}

\maketitle    

\begin{abstract}
Cosmic voids are large underdense regions that, together with galaxy clusters, filaments and walls, build up the large-scale structure of the Universe.
The void size function provides a powerful probe to test the cosmological framework.
However, to fully exploit this statistics, the void sample has to be properly cleaned from spurious objects. Furthermore, the bias of the mass tracers used to detect these regions has to be taken into account in the size function model.
In our work we test a cleaning algorithm and a new void size function model on a set of simulated dark matter halo catalogues, with different mass and redshift selections, to investigate the statistics of voids identified in a biased mass density field.
We then investigate how the density field tracers' bias affects the detected size of voids.
The main result of this analysis is a new model of the size function, parameterised in terms of the linear effective bias of the tracers used, which is straightforwardly inferred from the large-scale two-point correlation function.
This method is a crucial step in exploiting real surveys.
The proposed size function model has been accurately calibrated on halo catalogues, and used to validate the possibility to provide forecasts on the cosmological constraints, namely on the matter density contrast, $\Omega_{\rm M}$, and on the normalisation of the linear matter power spectrum, $\sigma_8$, at different redshifts.
\end{abstract}

\begin{keywords} 
large-scale structure -- cosmology:theory -- methods:statistical 
\end{keywords}


\section {Introduction}
Cosmic voids are large underdense regions from which matter is evacuated as a result of the collapse of the matter in between their boundaries and the repulsive action of dark energy (DE). They originate from the evolution of underdensities in the primordial density field. Voids constitute a major component of the Universe: while galaxy clusters enclose most of the mass, voids are the dominant spatial elements, accounting for about $90\%$ of the entire volume of the Universe \citep{2007MNRAS.380..551P}. Their sizes span over a wide range of scales, from diameters of few Mpc ({\em minivoids}) to about $200 \ \mbox{Mpc}$ ({\em supervoids}) \citep{Tikhonov, Szapudi}. Voids are only mildly non-linear objects, and tend to become more spherical as they evolve \citep{icke1984voids, vdW&kampen, Sheth}, which suggests that their isolated evolution should be easier to reconstruct than that of positive perturbations, despite their sphericity can be compromised during their growth and merging.

Thanks to their relatively simple structure and shape, voids represent the ideal environment to test a variety of cosmological parameters. They represent a population of statistically ideal spheres with a uniform distribution in a homogeneous and isotropic universe, so that their observed shape can be used to probe the assumed cosmological model by means of the Alcock-Paczy\'nski (AP) test \citep{AP1979} \citep[see e.g.][]{Lav&Wan2012, sutter2012, sutter2014, hamaus2014, Hamaus2016}. Moreover, being almost completely devoid of matter by definition, voids are extremely sensitive to diffuse components and have indeed been shown to possess great potential for constraining the DE properties \citep{2009Lee, Pisani}, in particular for scalar field DE models \citep{Bos2012, Adermann2017,Adermann2018}, and the mass of neutrinos \citep{massara2015, Kreisch2018, martinDE&neutrinos_prop}.
Thanks to their intrinsic low-density environment, cosmic voids have also proved to be promising objects to study modified gravity theories, since the effects of these scenarios, alternative to the General Relativity (GR), are expected to be more prominent in voids \citep{Clampitt2013, cai2015, Barreira2015, Zivick2015, Falck2018, Martin_modifiedG}. Deviations from GR can be observed also measuring the matter density profile of cosmic voids, which can be reconstructed exploiting voids as weak gravitational (anti-)lenses to infer their projected surface mass density \citep[see e.g.][]{Melchior2014, clampitt2015, sanchez2017,  Davies2018}. Modified gravity also causes a faster expansion around these objects, that can be revealed measuring the redshift-space distortions (RSD) in the cross-correlation of galaxies and void centres \citep{Hamaus2015, cai2016, Hamaus2017, Achitouv2017, Hawken}. One of the possible advantages of studying RSD around cosmic voids is that, in these regions, galaxy velocities are dominated by coherent bulk flows. Therefore the non-linear contributions can be in principle neglected and the RDS can be modelled using linear theory only \citep{Nadathur2019a,Natathur2019b}.

To exploit cosmic voids as cosmological probes, their statistical properties have to be modelled reliably \citep{NadathurI, NadathurII, Pollina2016}. In this work we focus on void abundances. The same excursion-set approach used for the mass function of dark matter (DM) haloes can be used also to model the size function of cosmic voids \citep{Sheth}. However, this model cannot accurately reproduce the number function of voids identified in cosmological simulations. Therefore, many studies have been conducted to better understand the evolution of voids over cosmic time and their statistics \citep{Jennings, Pisani, Achitouv2015, pycke&russel2016, Wojtak2016}. Moreover, the distribution of luminous tracers, such as e.g. galaxies and galaxy clusters, that are used to identify the voids, is biased with respect to the distribution of the underlying DM. It has been shown that the tracer bias plays a crucial role in determining the void profiles and size distributions. Having a reliable model to account for the effect of the tracer bias is thus mandatory to extract robust cosmological constraints from void statistics \citep{pollinalinear, pollinarelative, Natathur2019b}.

Recently, \citet{Roncarini2019} tested the void size function model developed by \citet{Sheth}, as revisited by \citet{Jennings}, on a series of unbiased simulated tracer catalogues, and extended the model to the case of voids identified in the distribution of DM haloes. In this work, we further validate the model on a larger set of 
catalogues with different mass and redshift selections. Moreover, we provide a new parameterisation of the void size function model as a function of the large-scale effective linear bias of the tracers. This represents a crucial ingredient to extract cosmological constraints from the statistical distribution of voids detected from real galaxy or cluster catalogues, when no direct information on the DM field is available. Finally, we investigate the cosmological constraints that can be inferred from the void size function at different redshifts.

Our work is organised as follows. In Section \ref{sec:method} we outline the methods employed for the identification of voids and the procedure of data reduction. In Section \ref{sec:void_SF} we present the theoretical definition of cosmic voids and some of the existing models developed for the void size function, then we describe the method adopted to rescale the abundances of voids identified in the mass tracer distribution as a function of the tracer bias.  
In Section \ref{sec:results} we apply the techniques previously described to simulated halo catalogues with different redshift and mass selections. We provide a relation between the effective linear bias of all the tracers and the one estimated inside voids, which is the one we use to rescale the void size function model. Then we measure the void size function in all our halo catalogues, and compare it to the new theoretical model, exploiting the void abundances to test the possibility of deriving constraints on the main cosmological parameters. Finally, in Section \ref{sec:conclusions}, we summarise our results and discuss future developments of this work.


\section {Void catalogues}
\label{sec:method}
In this Section, we present the set of $\Lambda$CDM N-body simulations used in our work, and the methods applied to build and clean the catalogues of cosmic voids.
With the cleaning algorithm we aim at aligning the objects included in the void catalogue with the adopted definition of cosmic void, which is fundamental to derive measured void size functions in agreement with theory predictions.

\subsection{Simulated halo catalogues}
In this work we make use of simulated halo catalogues extracted from a set of high resolution N-body simulations \citep[][]{baldi2012} of the standard $\Lambda$CDM cosmology, performed with the {\small C-GADGET} module \citep[][]{baldi2010}. We adopt a model consistent with WMAP7 constraints \citep{komatsu2011}, with $\sigma_8=0.809$, $h_0=0.703$, $\Omega_{\lambda}=0.7289$, $\Omega_{\rm M}=0.2711$, $\Omega_b=0.0451$, and a power spectrum with an initial scalar amplitude of $\mathcal{A}_s=2.194\cdot10^{-9}$ and a primordial spectral index of $n_s=0.96$.
The simulations followed the dynamical evolution of $2\cdot 1024^3$ particles: half of them are DM particles, while the other half is composed by non-collisional gas particles. Specifically, the catalogue covers a volume of $(1\, \text{Gpc}/h)^3$, with a mass resolution of $\sim 6 \cdot 10^{10}\;M_\odot/h$ for the DM particles. To test the procedure described in Section \ref{subsec:SF_rescaled}, we built a set of DM halo catalogues with a Friends-of-Friends (FoF) algorithm \footnote{The algorithm makes use of a linking length $\ell = 0.2 \cdot \overline{d}$, where $\overline{d}$ is the mean interparticle separation, gathering the CDM particles as primary tracers of the local mass density, and then attaching baryonic particles to the FoF group of their nearest neighbour.}, applying five different mass selection cuts: $\{2 \cdot 10^{12}\,, 2.5 \cdot 10^{12}\,, 5 \cdot 10^{12}\,, 7.5 \cdot 10^{12}\,, 10^{13}\,M_\odot /h\}$, at three different redshifts $\{z=0\,, 0.55\,, 1\}$. These mass cuts are applied to the  FoF mass in order to inspect a sufficiently wide range of values for tracers' bias \footnote{Other methods to measure halo masses were applicable in this case, e.g. using spherical overdensity masses. 
Anyway, the mass-cut criterion, as well as the redshift selection, are not relevant in our work and do not influence the outcomes of the manuscript.}.
The redshifts are chosen instead to span a significant fraction of cosmic time over which FoF haloes with masses greater than $10^{12} \ M_\odot /h$ are resolved. This range allows to test our methodology on haloes corresponding to common density peaks (low redshifts, low masses) and on newly forming haloes corresponding to rare density peaks.
The results obtained for the halo catalogues with $M_{\text{min}} = 2.5 $ and  $7.5 \cdot 10^{12} \ M_\odot /h$ are consistent with the ones of the other catalogues, and do not add any relevant information to the overall outcome of the paper. Thus, we will not show them in the Figures, with the only exception of Fig. \ref{fig:bias_z}.


\subsection{Building and cleaning the void catalogues}
\label{subsec:finder_cleaner}
Many different void finders have been developed over the last decades due to the non general concordance in the definition of voids \citep[see e.g.][]{Colberg_finders, micheletti_vimos, Elyiv2015}.
In this paper, we make use of the Void IDentification and Examination toolkit ({\small VIDE}) \citep{SutterVIDE} to construct our void catalogues. {\small VIDE} belongs to the class of algorithms based on geometrical criteria. It implements an enhanced version of the ZOnes Bordering On Voidness ({\small ZOBOV}) algorithm \citep{2008MNRAS.386.2101N}. {\small ZOBOV} is a popular publicly available code that finds density depressions in a three-dimensional set of points, without any free parameter or assumption about the void shape. The algorithm is based on a procedure called \textit{Voronoi tessellation}, which associates to each tracer a cell of volume that is closer to it than to any other tracer. Then the local density minima are found, and the \textit{watershed technique} is performed. Specifically, the shallower zones are merged together starting from the minima, forming a hierarchical tree of voids and subvoids. The process of rising the density threshold goes on until a deeper zone is encountered. The effective radius of voids is defined as the radius of a sphere containing the same volume as the watershed region, and the void centre is defined as the volume-weighted barycentre, $\overline{X}$, of the $N$ Voronoi cells that define the void,
\begin{equation}
\overline{X}  = \frac{\sum_{i=1}^{N}{\overline{x}_i V_i}}{\sum_{i=1}^{N}{V_i}} \textrm{ ,}
\end{equation}
where $\overline{x}_i$ are the coordinates of the $i$-th tracer of that void, and $V_i$ the volume of its associated Voronoi cell. Therefore the void centre does not necessarily coincide with the position of a tracer.

Once a candidate void catalogue is built, we apply the pipeline introduced in \citet{Ronconi2017}, which has been recently implemented in the {\small CosmoBolognaLib}\footnote{The {\small CosmoBolognaLib} \citep{CBL} is a large set of {\em free software} C++/Python libraries that provide an efficient numerical environment for cosmological investigations of the large-scale structure of the Universe. Thanks to the large amount of classes and functions recently implemented, these libraries offer the necessary tools to analyse cosmic void catalogues and perform all the statistical analyses of this work. The libraries are freely available at the following GitHub repository: \url{https://github.com/federicomarulli/CosmoBolognaLib}.}. The procedure standardises the outcome of void finders so as to make them directly comparable to model predictions. The cleaning algorithm is totally independent of the void finder employed since it makes use of the positions of void centres only. The goal is to take a candidate list of void centres and produce a catalogue of non-overlapping spherical underdensities, ``void''. Our cleaning algorithm can be divided in three main steps:

\begin{itemize}
\item The underdense regions that do not satisfy the following criteria are rejected from the catalogue: \textit{(i)} the effective radii have to be to greater than a given scale, $R_{\text{min}}$, which is chosen to remove objects that are under a certain resolution threshold; \textit{(ii)} the central density has to be lower than {$(1+\delta_v^{NL}) \overline{\rho}$}, where $\delta_v^{NL}$ is a given non-linear underdensity threshold (see Section \ref{subsec:SF_rescaled}), and  $\overline{\rho}$ is the mean density of the tracers. In this way we are rejecting voids that are not relevant for our analysis, that is those regions that cannot be defined as cosmic voids according to our definition.
\item The effective void radii are rescaled: the algorithm reconstructs the density profile of each void and the value of the radius is increased until the sphere reaches a specific density contrast threshold, $\delta_v^{NL}$. This value is not universal, any other threshold sufficiently high to enclose enough tracers and sufficiently low to identify the voids would be valid (this will be intensively discussed in Section \ref{subsec:SF_rescaled}).
\item When two voids do overlap (thus when the distance between void centres is less than the sum of their radii), the one with the higher central density is rejected, avoiding double countings. This choice favours the selection of larger, most underdense voids.
\end{itemize}

The effect of the cleaning procedure is to reshape the selected voids as spherical non-overlapping regions, centred in density depths of the tracer density field, embedding a fixed density contrast (see Section \ref{sec:void_SF}). 
As a consequence, the void number counts result lower with respect to the original output of {\small VIDE}. Moreover, as it can be seen from Table \ref{tab:counts}, \textit{(i)} the total number of void counts tends to decrease for tracer catalogues with higher mass selections due to the lowering of the resolution, and \textit{(ii)} the void radii are shifted towards higher values because of the consequent reduction of the mean mass density.

Our choice of modelling the underdensity regions as spheres is aimed at comparing void statistics directly to theoretical models. We do not need to reconstruct accurately the real shape of individual voids. Although real voids are not spherical objects, the mean void ellipticity is small in standard cosmological frameworks \citep{verza2019}. We can thus reasonably assume that the voids' geometry is spherical on average \citep{Lav&Wan2012}.

\begin{table}
\centering
\caption{Void counts in 5 logarithmic bins of void effective radii, $\overline{R}_\textrm{eff}$, in the range $[18$-$60] \ \text{Mpc}/h$, for DM halo catalogues with different mass and redshift selections, after the cleaning procedure has been applied.} \label{tab:counts}

\begin{tabular}{ccccccc}

\hline
\toprule
\rowcolor{Gray}
\multicolumn{7}{c}{$z=0.00$}  \\ 
\hline \noalign{\smallskip}
\multirow{2}{*}{} & \multirow{2}{*}{} & \multicolumn{5}{c}{$\overline{R}_\textrm{eff}$ $[{\text{Mpc}/h}]$} \\
\cline{3-7} \noalign{\smallskip}
& & $20.5$ & $26.0$ & $33.1$ & $42.1$ & $53.6$ \\
{$M_\textrm{min}$ $[{M_\odot /}{h}]$} & {$N_\textrm{tot}$} & \multicolumn{5}{c}{$N(\overline{R}_\textrm{eff})$} \\
\midrule
$2\cdot10^{12}$    & $1063$ & $719$ & $288$ & $53$  & $3$   & $0$  \\
$5\cdot10^{12}$    & $1007$ & $544$ & $333$ & $115$ & $15$  & $0$  \\
$10^{13}$          & $803$  & $291$ & $309$ & $160$ & $39$  & $4$  \\
\noalign{\smallskip}
\hline

\toprule
\rowcolor{Gray}
\multicolumn{7}{c}{$z=0.55$}  \\
\hline \noalign{\smallskip}
\multirow{2}{*}{} & \multirow{2}{*}{} & \multicolumn{5}{c}{$\overline{R}_\textrm{eff}$ $[{\text{Mpc}/h}]$} \\
\cline{3-7} \noalign{\smallskip}
& & $20.5$ & $26.0$ & $33.1$ & $42.1$ & $53.6$ \\
{$M_\textrm{min}$ $[{M_\odot /}{h}]$} & {$N_\textrm{tot}$} & \multicolumn{5}{c}{$N(\overline{R}_\textrm{eff})$} \\
\midrule
$2\cdot10^{12}$    & $1053$ & $690$ & $301$ & $56$  & $6$   & $0$  \\
$5\cdot10^{12}$    & $943$  & $444$ & $356$ & $120$ & $22$  & $1$  \\
$10^{13}$          & $693$  & $196$ & $256$ & $176$ & $49$  & $7$  \\
\noalign{\smallskip}
\hline

\toprule
\rowcolor{Gray}
\multicolumn{7}{c}{$z=1.00$}  \\
\hline \noalign{\smallskip}
\multirow{2}{*}{} & \multirow{2}{*}{} & \multicolumn{5}{c}{$\overline{R}_\textrm{eff}$ $[{\text{Mpc}/h}]$} \\
\cline{3-7} \noalign{\smallskip}
& & $20.5$ & $26.0$ & $33.1$ & $42.1$ & $53.6$ \\
{$M_\textrm{min}$ $[{M_\odot /}{h}]$} & {$N_\textrm{tot}$} & \multicolumn{5}{c}{$N(\overline{R}_\textrm{eff})$} \\
\midrule
$2\cdot10^{12}$    & $1090$ & $698$ & $314$ & $72$  & $6$   & $0$  \\
$5\cdot10^{12}$    & $850$  & $370$ & $301$ & $146$ & $33$  & $0$  \\
$10^{13}$          & $557$  & $140$ & $170$ & $156$ & $77$  & $14$  \\
\hline
\bottomrule

\end{tabular}

\end{table}


\begin{figure*}
\plotfull{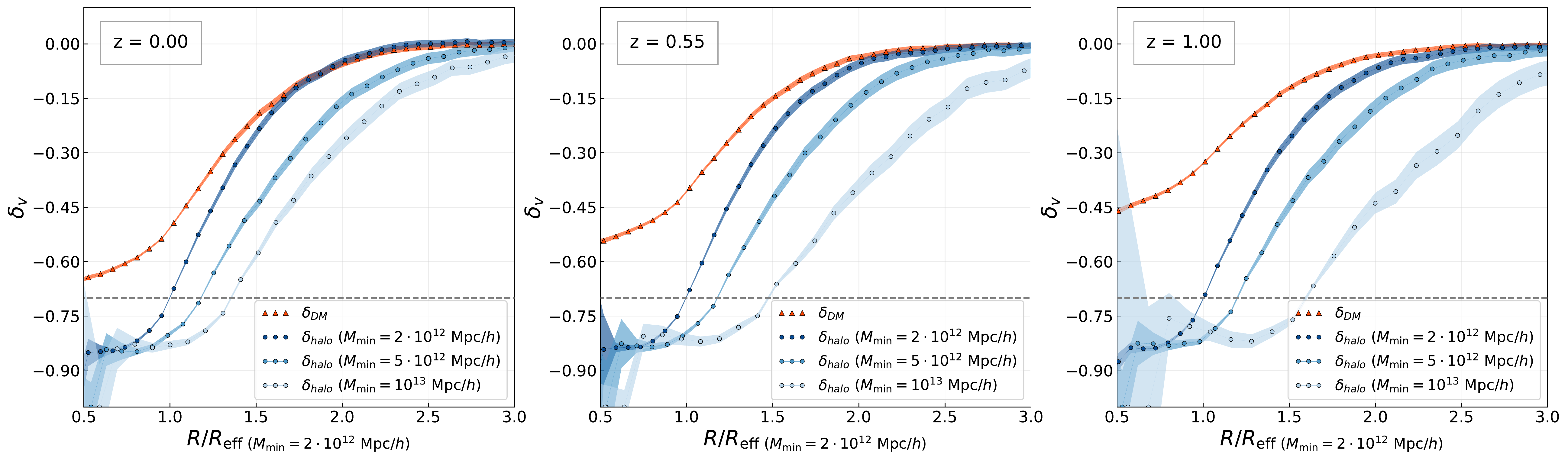}
    \caption{Spherically-averaged density profiles measured from the centres of voids identified in the tracer distribution at redshifts $z=0$ (\textit{left}), $z=0.55$ (\textit{centre}), $z=1$ (\textit{right}). The red lines represent the median of the profiles computed in the DM particle distribution, while the blue ones indicate the profiles in the DM halo catalogues with different mass-cuts. The horizontal dashed line indicates the value of the density contrast threshold ($\delta_{v,\, \text{tr}}^{NL} = -0.7$) selected in the cleaning procedure. All the profile radii are rescaled to the mean effective radius of the catalogue with $M_\textrm{min} = 2 \cdot 10^{12} \ {M_\odot /}{h}$, in order to show the effect of the rescaling procedure of the cleaning algorithm. The shaded areas represent $2\sigma$ confidence regions, that is 2 times the standard deviation of the distribution of the mean values.}
    \label{fig:density_profiles}
\end{figure*}


\section {The void size function model}
In this Section we first present the theoretical model for the void size function that we use in our work. Secondly, we describe the method we adopt to rescale the model to make it directly comparable with the abundance of voids identified in the distribution of biased tracers. In the end, we focus on the measure of the tracer bias in overdensity and underdensity regions, since the value of the latter is required for the re-parameterisation of the theoretical model.
\label{sec:void_SF}

\subsection{The size function model of voids detected in the DM
distribution}
\label{subsec:void_SF}

Contrary to what happens in the case of overdensities, voids typically do not invert their expansion during their growth, so they cannot collapse and virialise like DM haloes. Instead, they expand at a {\em super-Hubble} rate, which is inversely proportional to the density enclosed in their boundary. Considering an initial negative top-hat perturbation, and modelling it as a set of concentric shells, the inner shells will expand faster than the outer ones. This implies that the shells near the centre of the underdensity will eventually reach the more external ones. This event is called {\em shell-crossing}. When this occurs, we can consider that a void is formed. After the shell-crossing, the void recovers the overall expansion rate, growing with the Hubble flow. This phenomenon is completely described by the spherical non-linear evolution model of an isolated spherically-symmetric density perturbation. It can be demonstrated that in linear theory this event takes place at a fixed value of the density contrast, $\delta_v^{NL}\approx -2.71$, for an Einstein-de Sitter universe (EdS). Therefore, we can define voids as underdense, spherical, non-overlapping regions, which have gone through shell-crossing. 
 
The void size function, that is the comoving number density of cosmic voids as a function of their effective radii, has been modelled for the first time by \citet{Sheth} (hereafter the SvdW model), with the same excursion-set approach used to model the mass function of DM haloes \citep{P&S, Bond}. The void size function in linear theory can be written as follows:
 \begin{equation}
  \label{eq:vsf02}
  \frac{\mathrm{d}n}{\mathrm{d}\ln r}\biggr|_{\text{lin}} = \frac{f_{\ln\sigma} (\sigma)}{V(r)} \frac{\mathrm{d}\,\ln\sigma^{-1}}{\mathrm{d}\,\ln r}\ \text{,}
\end{equation}
 where $f_{\ln\sigma}$ is the fraction of the Universe occupied by cosmic voids, as predicted by the excursion-set theory:
\begin{equation}
  \label{eq:SF1}
  f_{\ln\sigma} = 2 \sum_{j=1}^{\infty}j \pi x^2 \sin(j \pi
  \mathcal{D})\exp\biggl[-\frac{(j \pi x)^2}{2}\biggr]\, ,
\end{equation}
where
\begin{equation}
  \label{eq:SF2}
  x \equiv \frac{\mathcal{D}}{|\delta_v^L|}\,\sigma \, ,
\end{equation}
and
\begin{equation}
  \label{eq:SF3}
  \mathcal{D} \equiv \frac{|\delta_v^L|}{\delta_c^L + |\delta_v^L|}\, .
\end{equation}
In the previous equations, $\sigma$ is the square root of the mass variance, while $\delta_v^L$ and $\delta_c^L$ represent the shell-crossing threshold and the critical value for the collapse of an overdense shell in an EdS universe, respectively\footnote{In this paper we indicate with the superscripts \textit{L} and \textit{NL} the density contrasts derived in linear and non-linear regime, respectively. In absence of any superscript, we take for granted the reference to the non-linear counterpart. Moreover, with the subscript $v$ we refer to the values measured inside voids, both for DM and biased mass tracers. We will use the subscript \textit{tr} to indicate generically any type of mass tracers, and the subscript \textit{halo} to indicate specifically the DM haloes.}. The latter is expected to vary within $1.06 \leq \delta_c^L \leq 1.686$, since both the turn-around and the collapse density contrast value can be considered acceptable assumptions.

In order to derive the void size function in the non-linear regime, SvdW assumed that the total number of voids has to be conserved in the transition from linearity to non-linearity. This condition leads to a correction in the void radius by a factor $C \propto (1 + \delta_v^{NL})^{-1/3}$:
\begin{equation}
  \label{eq:vsf03}
    \frac{\mathrm{d}\,n}{\mathrm{d}\, \ln r} \biggr|_{\text{SvdW}}
    = \frac{\mathrm{d}\,n}{\mathrm{d}\, \ln (C\,r)}\biggr|_{\text{lin}}\ \text{.}  
\end{equation}
However, according to Eq. \eqref{eq:vsf03}, the fraction of the volume occupied by voids can be larger than the total volume of the Universe. To address this issue, \citet{Jennings} proposed a {\em volume conserving} model (hereafter the Vdn model), in which the total volume occupied by cosmic voids is conserved in the transition to the non-linear regime. In particular, the Vdn model can be obtained as follows:
\begin{equation}
  \label{eq:vsf04}
  \frac{\mathrm{d}\,n}{\mathrm{d}\, \ln r}\biggr|_{\text{Vdn}} =
  \frac{\mathrm{d}\,n}{\mathrm{d}\, \ln r}\biggr|_{\text{lin}} \frac{V(r_L)}{V(r)}
  \frac{\mathrm{d}\, \ln r_L}{\mathrm{d}\, \ln r}\ \text{,}
\end{equation}
where the subscript \textit{L} indicates a value derived in linear theory.
\citet{Roncarini2019} showed that the Vdn model can predict accurately the measured void size function of unbiased tracers, provided that the void catalogue is appropriately cleaned from spurious voids and the void radii are rescaled to a fixed density threshold (see Section
\ref{subsec:finder_cleaner}).

\begin{figure*}
\plottwo{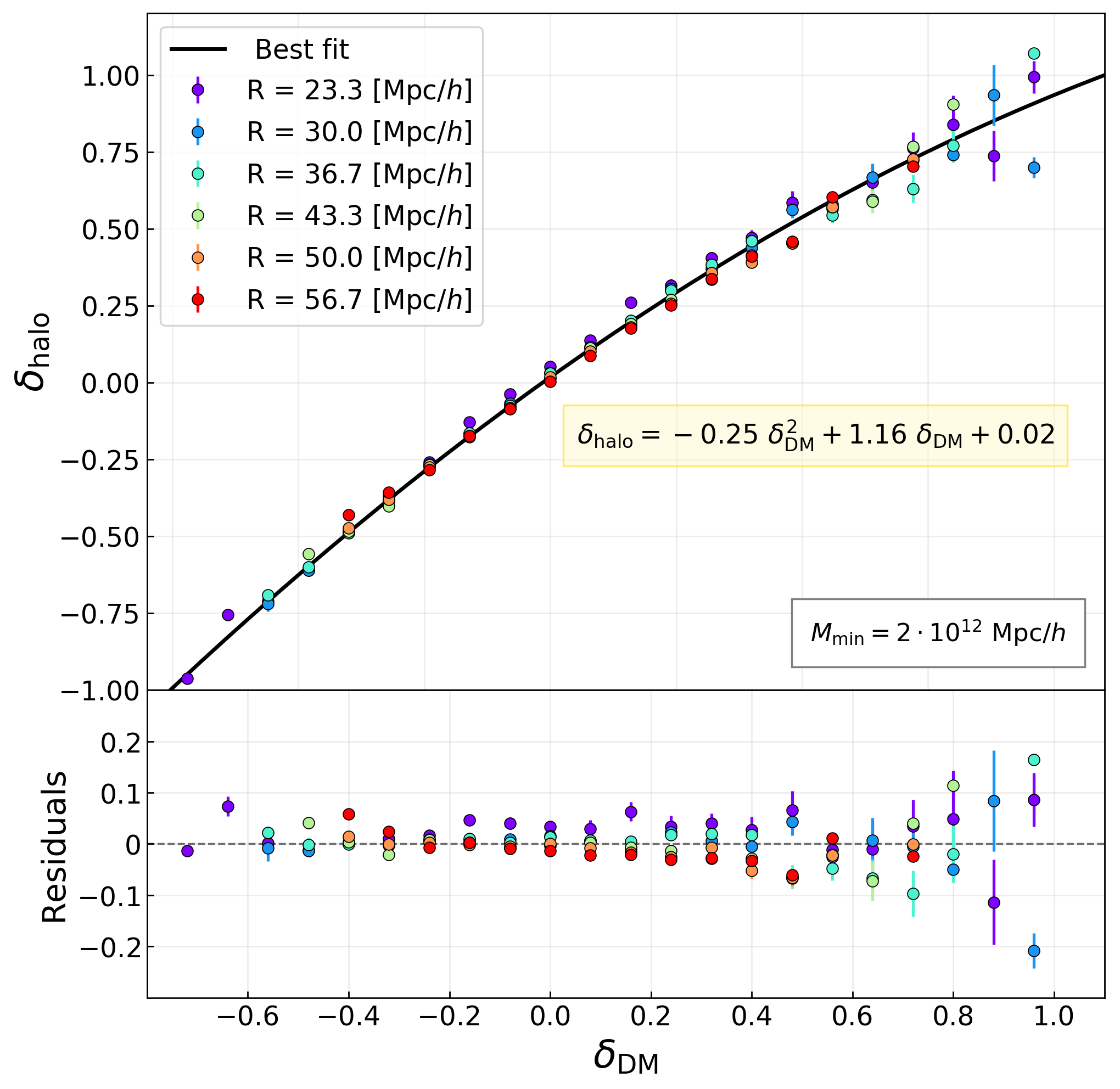}{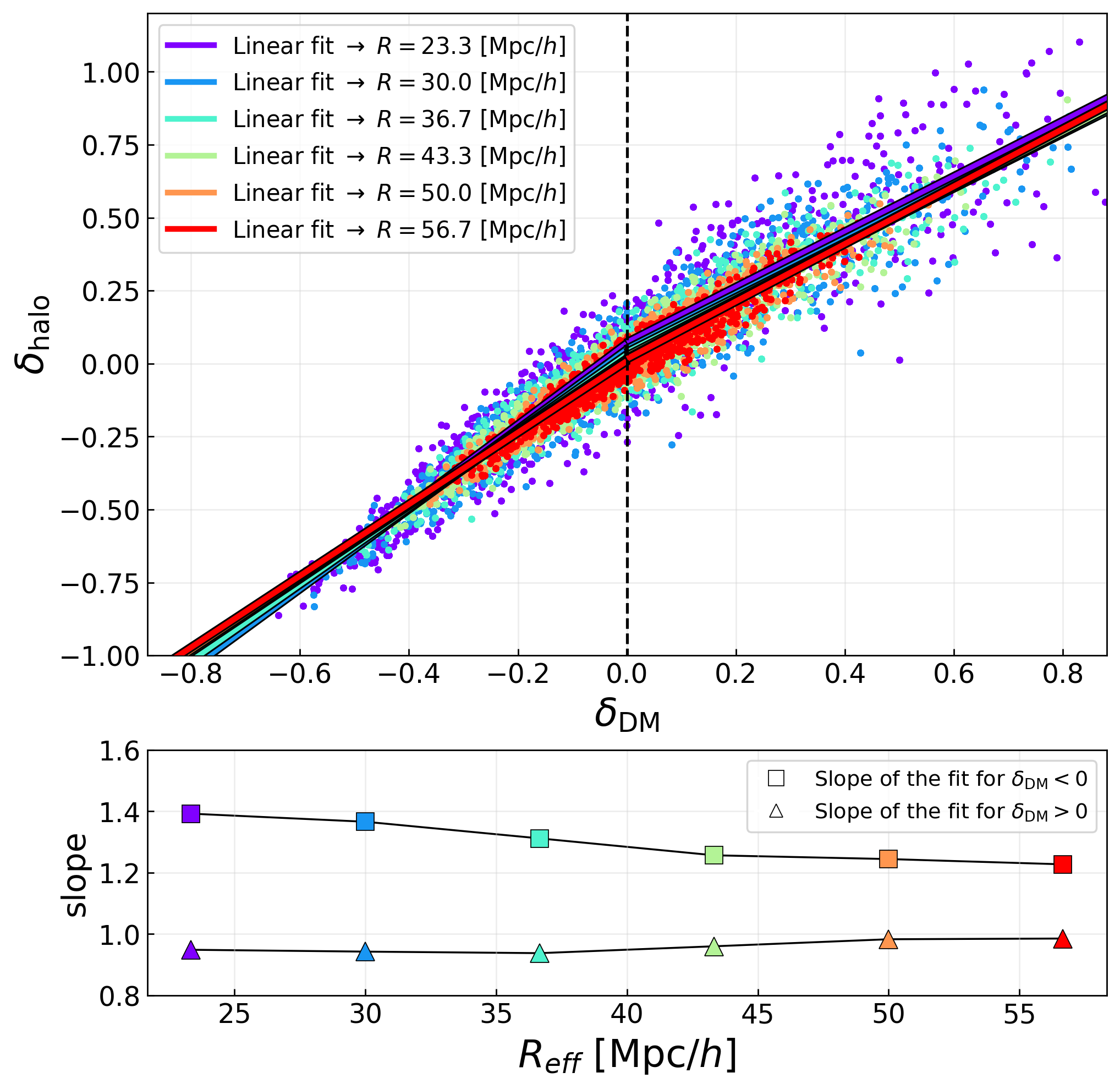}
    \caption{Relation between the density contrast computed in the DM distribution ($\delta_{\text{DM}}$) and in the tracer distribution ($\delta_{\text{halo}}$). The data points are computed as the spherically-averaged density contrast for $1000$ random positions in the halo catalogue with $M_\textrm{min} = 2 \cdot 10^{12} \ {M_\odot /}{h}$ at $z=0$, averaging in different radius bins. The different colours refer to the different radius sizes of the spheres used to compute the density contrast.
    \textit{Left}: in the \textit{upper sub-panel} the data are computed as the median of the values of $\delta_{\text{halo}}$ in different bins of $\delta_{\text{DM}}$, with error bars computed as the ratio between the standard deviation and the square root of the number counts in each bin. The points are fitted with a second-order polynomial, whose equation is reported in the yellow insert and represents the non-linear bias function. In the \textit{lower sub-panel} are reported the residuals from the quadratic fit. 
    \textit{Right}: in the \textit{upper sub-panel} the points with $\delta_{\text{DM}}>0$ and $\delta_{\text{DM}}<0$ of each radius bin are fitted separately with a linear relation. In the \textit{lower sub-panel} is shown the variation of the slope of each fit as a function of the radius of the sphere used to compute the averaged density contrast.}
    \label{fig:bias_compare}
\end{figure*}

\subsection{The size function model of voids detected in biased tracer distribution}
\label{subsec:SF_rescaled}
The goal of the cleaning procedure is to make the measured void size function directly comparable to the Vdn model. The method is based on the requirement that the spherically-averaged density contrast embedded inside the void effective radius has to coincide with the value defined by the theory. As shown in \citet{Roncarini2019}, dealing with unbiased tracers, this value can correspond to that of the shell-crossing in the non-linear regime, $\delta_{sc}^{NL} = -0.795$. With this prescription, the measured size function of voids identified in the DM field is consistent with the one predicted by the Vdn model. 

It is important to notice that the choice to rescale the void radii to the specific density contrast characteristic of the shell-crossing is not universal. In fact, in order to have correspondence with the theory, it is only required to rescale the radii at a chosen density threshold with the cleaning algorithm and use the same density contrast (converted to linear theory) also in the theoretical void size function. 
This overdensity threshold $\delta_{sc}^{NL}$, which identifies the time at which cosmic voids form, has to be rescaled at redshift $z>0$, using the growth factor:
\begin{equation}
    \label{eq:unbiased01}
    \delta_{sc}^{L}(z) = \delta_{sc}^{L}(0)\dfrac{\textit{D}(z)}{\textit{D}(0)} \, \text{,}
\end{equation}
where $\delta_{sc}^L(0) = -2.71$ is the shell-crossing density contrast in linear theory at $z=0$, and $\textit{D}(z)$ is the growth factor. Therefore, voids formed at lower density contrast values in the past.
After shell-crossing, the void radii continue to grow with cosmic time, and the enclosed volume becomes increasingly underdense. Choosing the threshold $\delta_v^{NL} = \delta_{sc}^{NL}(z)$, we are rescaling voids to the radii embedding the density contrast typical of the phenomenon of the shell-crossing at that epoch. In other words, we are rescaling void radii to the size they had when they formed. 
This method cleans the catalogue of newly forming voids that have yet to meet the underdensity criterion. In fact, the ones that cannot be rescaled to this particular density contrast (because they are not enough underdense at that epoch) are rejected. However, every negative density threshold $-1<\delta_v^{NL}<0$ is allowed in principle, provided that the same value is used in the theoretical size function. 

Dealing with mass tracers, the effect of the tracer bias has to be taken into account to extract accurate cosmological constraints from the void number counts \citep[see e.g.][]{pollinalinear}. Let us assume that the voids identified in the DM and in the mass tracer density fields have the same radii when the phenomenon of the shell-crossing occurs. This implies that voids found in the biased tracer distribution have a lower embedded density contrast with respect to the ones traced by DM particles. This is illustrated in Fig. \ref{fig:density_profiles}, that shows the spherically-averaged void density profiles\footnote{All the void profiles with effective radii, $R_{\text{eff}}$, larger than two times the mean inter-particle separation are stacked in these plots. As a result of the cleaning procedure, that rescales every void radii at the same level of density contrast, the profiles do not show a clear dependence on the void effective radius and they are therefore averaged together.} as traced by either DM or DM haloes with different biases, at three different redshifts \citep[see also][]{Roncarini2019}. With this assumption, and given that the DM density field within voids is linearly related to the density field traced by biased tracers \citep{pollinarelative}, the threshold at which the void radii have to be rescaled corresponds to:
\begin{equation}
  \delta_{v,\, \text{tr}}^{NL} = b\, \delta_{v,\, \text{DM}}^{NL}\ \text{,  with \ }  \delta_{v,\,\text{DM}}^{NL}=-0.795 \, \text{.}
\end{equation}
It is evident that for $b>1$ the density contrast can reach values so low that the phenomenon of the shell-crossing might not even happen, since the lowest minimum is $\delta_{v,\, \text{tr}}^{NL}=-1$ (corresponding to the state without any tracer). Therefore, it is not possible to perform this technique to rescale the void radii in the case of biased tracers. For this reason, we use a different density contrast in the rescaling procedure, fixing the threshold in the tracer distribution, instead of in the DM one. In particular, we set a threshold equal to $-0.7$ for all the considered halo catalogues. Thus we rescale all the voids found in the tracer catalogues to an effective radius such that the spherically-averaged density contrast they contain is $\delta_{v,\, \text{tr}}^{NL} = -0.7$. This choice is aimed at having not too small void radii (the higher is the threshold, the higher is the radius), in order to enclose a sufficient number of tracers. In fact, the resolution of the tracer catalogue does not allow us to identify voids with radii smaller than $2$-$3$ times the mean inter-particle separation. At the same time, we require that the chosen threshold is not too high, since the selected regions have to be enough underdense to be still classified as voids. Moreover, we want to keep the threshold low also to prevent the possible overlap between adjacent voids, that would make the cleaning algorithm to discard the smaller, thus the one with higher central density, to prevent double counts (see Section \ref{subsec:finder_cleaner}).

To model the theoretical size function of voids identified in the mass tracer field, we follow the prescription described in \citet{Roncarini2019}. This is based on the reasonable assumption that voids identified in the DM and in the tracer field are equal in number, and that their centre positions are approximately the same\footnote{We tested this hypothesis using a catalogue of voids identified in the DM density field and cleaning it using the corresponding distribution of DM haloes as tracer. The results obtained are in agreement with the ones found with the voids identified in the biased tracer distribution. Therefore this assumption can be considered statistically valid, even if the correspondence between void centres in different mass density fields is not always exact.}. Since the Vdn model can predict the number of voids with a certain radius, the simplest procedure to apply is to rescale the theoretical size function dividing the chosen threshold by the bias value:
\begin{equation}
  \delta_{v,\, \text{DM}}^{NL}= \frac{\delta_{v,\, \text{tr}}^{NL}}{b} \ \text{,  with \ }  \delta_{v,\,\text{tr}}^{NL}=-0.7 \, \text{.}
\end{equation}
We convert $\delta_{v,\, \text{DM}}^{NL}$ to its linear counterpart, with the fitting formula provided by \citet{bernardeau}: 
\begin{equation}
  \label{eq:bias02}
  \delta_{v,\, \text{DM}}^L = \mathcal{C}\, \bigl[1 - (1 + \delta_{v,\, \text{DM}}^{NL})^{-1/\mathcal{C}} \bigr]\, ,
\end{equation}
with $\mathcal{C}=1.594$. This equation is exact for models with null cosmological constant $\Lambda$, and is a good fit for any values of $\Lambda$, especially for the underdense regions.
The density contrast given by Eq. \eqref{eq:bias02}, $\delta_{v,\, \text{DM}}^L$, has to be used in Eqs. \eqref{eq:SF2} and \eqref{eq:SF3}.
This is basically equivalent to expand the radii of voids identified in the DM field (embedding the same density contrast $-0.7$), in order to match the same radius of the ones identified in the tracer field. In this way, we are able to recover the theoretical size function taking into account the effect of the bias, that shifts the size function to higher void radii. \footnote{A new algorithm to rescale the void size function model as a function of the tracer bias has been implemented in the {\small CosmoBolognaLib}. The code requires in input the values of the radii at which the model is computed, the redshift of the sample, the size function model to use (e.g. SvdW, Vdn) and the effective bias of the catalogue, $b_{\text{eff}}$. The latter can be automatically converted to $b_{\text{rel}}$ using the relation calibrated in this work (see Section \ref{subsec:relation_bias}).
Moreover, a new notebook is provided to explain, step by step, how to clean a void catalogue, and how to measure and model the void size function, according to the method described in this paper.}


\begin{figure*}
\plotfull{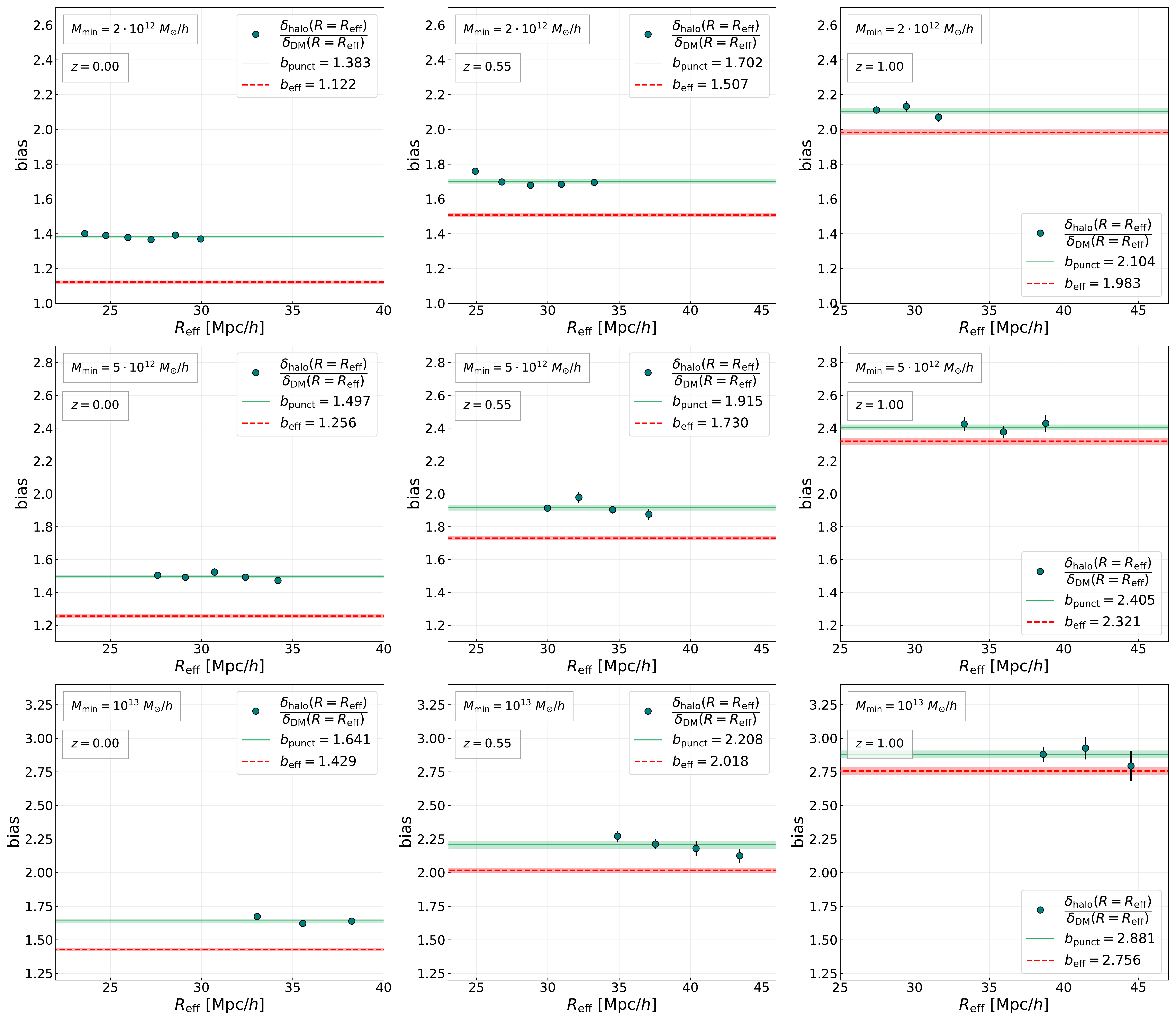}
    \caption{Measure of the tracer bias estimated as the ratio between the density contrast computed in the halo ($\delta_{\text{halo}}$) and in the DM ($\delta_{\text{DM}}$) density fields, at a distance of $1 \ R_{\text{eff}}$ from the void centres. The different panels show the results obtained from the halo catalogues with $M_\textrm{min} = 2 \cdot 10^{12} \ {M_\odot /}{h}$, $5 \cdot 10^{12} \ {M_\odot /}{h}$ and $10^{13} \ {M_\odot /}{h}$ (rows from \textit{top} to \textit{bottom}), at redshifts $z=0$, $z=0.55$, $z=1$ (columns from \textit{left} to \textit{right}). The dark green points represent the median of the ratio for different radius bins, with error bars representing the $1 \sigma$ uncertainty. The green lines are the weighted fit of the data, $b_{\text{punct}}$, while the red dashed lines show the effective bias, $b_{\text{eff}}$. The shaded regions show the $1 \sigma$ errors on the bias values.}
    \label{fig:bias_R}
\end{figure*}

\subsection{The bias of tracers in overdensity and underdensity regions}
\label{subsec:bias_void}
The bias of cosmic tracers is a non-linear stochastic function described by the conditional probability of tracer density contrast, $\delta_{\text{tr}}$, given the mass density contrast $\delta_{\text{DM}}$ \citep[see e.g.][and references therein]{Dekel1999, DiPorto2016}. This is shown in Fig. \ref{fig:bias_compare}, where the density contrast of a 
halo catalogue analysed in this work ($M_\textrm{min} = 2 \cdot 10^{12} \ {M_\odot /}{h}$ at $z=0$) is plotted against the corresponding DM density contrast, smoothing the density field at $1000$ random positions with top-hat spherical filters with different radii. 
As shown in the left panel of Fig. \ref{fig:bias_compare}, the data are well fitted by a second-order polynomial. However, a linear model is accurate enough to describe separately the points in the overdensity and in the underdensity regions. Indeed, fitting all the points with a second-order polynomial the reduced chi square is $\tilde{\chi}^2= 1.977$, while fitting $\delta_{\text{DM}}>0$ and $\delta_{\text{DM}}<0$ separately with a linear relation we obtain $\tilde{\chi}^2= 1.758$ and $\tilde{\chi}^2= 2.780$, respectively. The slope of the former, $b \ [\delta_{\text{DM}}>0]$, represents the linear bias that can be approximately inferred e.g. from the tracer large-scale two-point correlation function (2PCF). The slope of the latter, $b \ [\delta_{\text{DM}}<0]$, represents the bias of the tracers inside cosmic voids, which is the value we actually need in order to properly rescale the void size function, as we will explain in the next Section. As shown in the right panel of Fig. \ref{fig:bias_compare}, $b \ [\delta_{\text{DM}}<0]>b \ [\delta_{\text{DM}}>0]$.

Since $b \ [\delta_{\text{DM}}<0]$ is generally not directly measurable, we shall calibrate a relation between $b \ [\delta_{\text{DM}}>0]$ and $b \ [\delta_{\text{DM}}<0]$ to be able to model the size function of voids detected from real tracer catalogues. Specifically, we search for a relation between the effective linear bias of the tracers used to detect cosmic voids, $b_{\text{eff}}\sim b \ [\delta_{\text{DM}}>0]$ (that we measure from the tracer 2PCF at large scales, as described in Appendix \ref{Appendix_A}), and the linear bias of tracers {\em inside} the detected voids. A convenient estimate of the latter can be assessed through the ratio between $\delta_{v,\, \text{halo}}^{NL}$ and $\delta_{v,\, \text{DM}}^{NL}$ at a distance of $R_{\text{eff}}$ from void centres \citep{Roncarini2019}:
\begin{equation}
\label{eq:b_punct}
b_{\text{punct}} \equiv \bigg \langle \frac{\delta_{v,\, \text{tr}}^{NL}(R=R_{\text{eff}})}{\delta_{v,\, \text{DM}}^{NL}(R=R_{\text{eff}})} \bigg \rangle \, .
\end{equation}
The {\em punctual bias} given by Eq. \ref{eq:b_punct} characterises the relation between the density contrast measured in the tracer field and in the DM field {\em punctually}, that is at $R=R_{\text{eff}}$. 
Since in our analysis the value of $\delta_\text{halo}(R_{\text{eff}})$ is fixed at $-0.7$, then $\delta_{\text{DM}}(R_{\text{eff}})$ is exactly the value we need to rescale the void size function model (see Section \ref{subsec:SF_rescaled}).

An alternative method to estimate $b \ [\delta_{\text{DM}}<0]$ is the one employed by \citet{pollinarelative}. They found a linear relation between the density profiles of tracers and DM inside voids. The slope of this relation, $b_{\text{slope}}$, provides an estimate of the tracer bias in underdensity regions. We discuss about this method in Appendix \ref{Appendix_B}, but choose not to use it in our analysis as it is more prone to uncertainties.


\section {Results}
\label{sec:results}
In this Section, we first estimate the linear bias of DM haloes inside voids, $b_{\text{punct}}$.
We then model the relation between this bias and the effective linear bias of all the tracers used to detect the voids, $b_{\text{eff}}$.
The latter is estimated from the tracer 2PCF at large scales, as explained in Appendix \ref{Appendix_A}.
We want a size function model that can be promptly compared with real data measures. To this end, it is crucial to obtain a relation between $b_\text{eff}$ and $b_\text{punct}$ that can be applied, independently of the tracer used to sample the underlying DM density field.
Afterwards, we measure the void size function in all our simulated catalogues, and compare the measurements with the theoretical model.
Finally, we perform a Bayesian Markov chain Monte Carlo (MCMC) statistical analysis, exploiting the calibrated bias scaling relation to construct the likelihood function. With this approach, we investigate the constraining power of the method by assessing the posterior probability of two cosmological parameters, namely $\Omega_{\rm M}$ and $\sigma_8$, at varying redshift.

\begin{figure}
\plotone{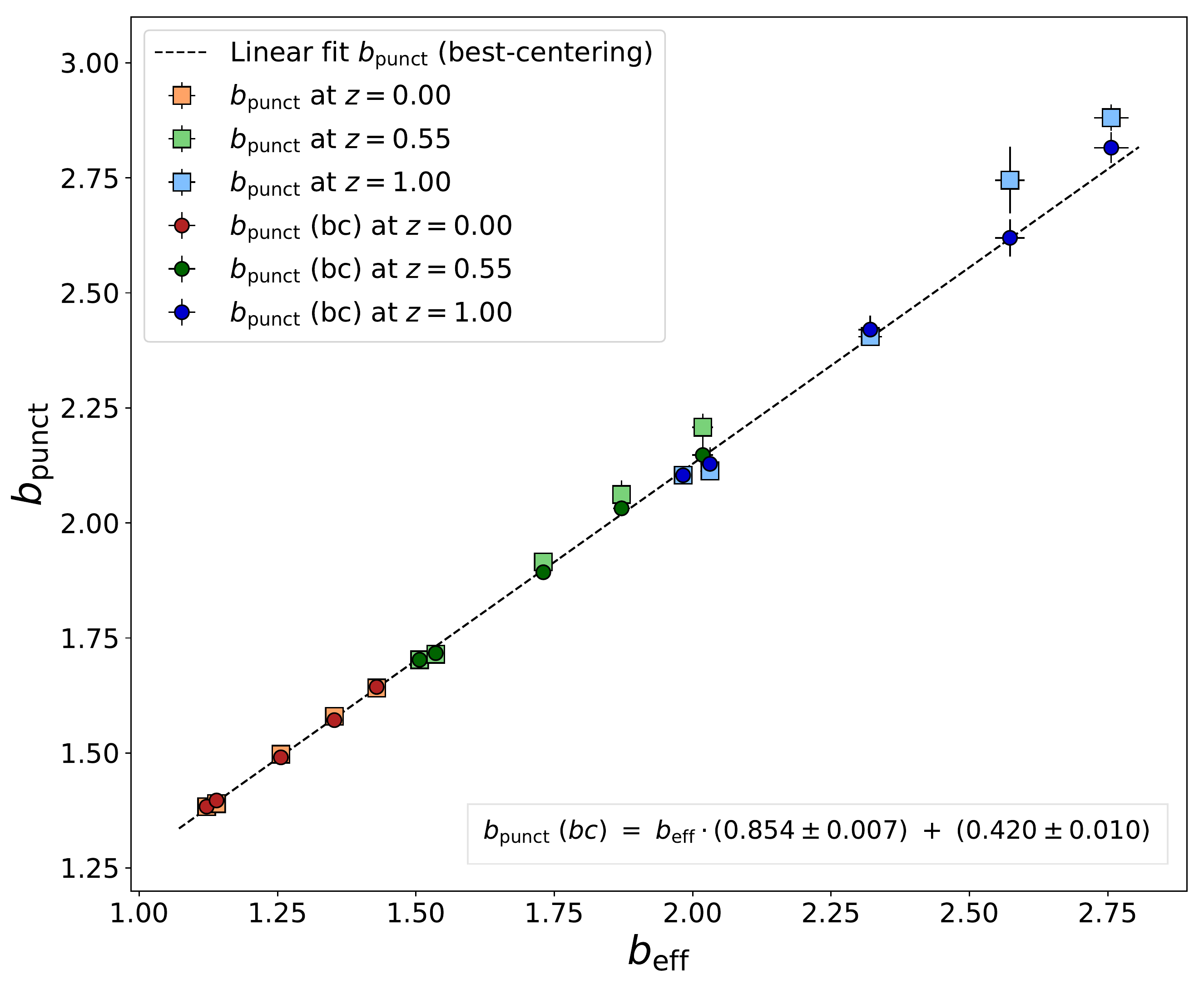}
    \caption{Relation between the effective bias ($b_{\text{eff}}$) and the bias measured inside voids ($b_{\text{punct}}$), at different redshifts. The points correspond to the data reported in Table \ref{tab:bias}, with $1 \sigma$ errors. The squares are the values of $b_{\text{punct}}$, obtained
    with the method presented in Fig. \ref{fig:bias_R}, while the circles are estimated with the \textit{best-centering} technique (see Section \ref{subsec:relation_bias}) and correspond to $b_{\text{punct}}\text{(bc)}$. The black line is the linear fit of the $b_{\text{punct}}\text{(bc)}$ values. The best-fit parameters are reported in the label in the lower right corner.}
    \label{fig:bias_z}
\end{figure}

\subsection{The bias of DM haloes inside voids}
\label{subsec:relation_bias}
Figure \ref{fig:bias_R} shows the ratio between the density contrast of haloes and DM, $\delta_{\text{halo}}/\delta_{\text{DM}}$, measured at $R=R_{\text{eff}}$ and averaged over voids of similar effective radii, together with their weighted average values (Eq. \eqref{eq:b_punct}), $b_{\text{punct}}$, for all the considered simulated catalogues. 
Therefore, in this Figure the points are  obtained by computing the ratio ${\delta_{v,\, \text{tr}}^{NL}(R=R_{\text{eff}})/\delta_{v,\, \text{DM}}^{NL}(R=R_{\text{eff}})}$ for each void of the catalogues (with $R_{\text{eff}}$ being the effective radius of that specific void) and binning the result as a function of $R_{\text{eff}}$. Then, to compute the value of $b_{\text{punct}}$, we perform a weighted fit of these data with a constant.
For comparison, we show also the effective tracer bias, $b_{\text{eff}}$, estimated from the 2PCF at large scales, as explained in Appendix \ref{Appendix_A}. As shown in Fig. \ref{fig:bias_R}, the $\delta_{\text{halo}}/\delta_{\text{DM}}$ ratio decreases as a function of $R_{\text{eff}}$, especially at high redshifts. In particular, it tends to $b_{\text{eff}}$ at large radii, in agreement with the results obtained by \citet{pollinalinear, pollinarelative}. Nevertheless, we find that an average constant value of $b_{\text{punct}}$ is sufficient to properly rescale the void size function, as we will show in Section \ref{subsec:data-theory}.

Since in most cases it is not possible to infer the underlying DM distribution inside voids, it is worth to search for a relation between $b_{\text{punct}}$ and $b_{\text{eff}}$, which can be accurately estimated e.g. from clustering measurements.
This relation is displayed in Fig. \ref{fig:bias_z}. As it can be seen, the data can be well fitted by a simple linear model. 

\begin{table}
\centering
\caption{The values of the bias with $1\sigma$ uncertainties measured in the overdensity ($b_{\text{eff}}$) and in the underdensity ($b_{\text{punct}}$ and $b_{\text{punct}}$ (bc)) regions, for all the halo catalogues with different mass selections and redshifts.} \label{tab:bias}
\begin{tabular}{ccccc}
\toprule
\multirow{2}{*}{$M_\textrm{min}$ [${M_\odot /}{h}]$} & \multicolumn{3}{c}{$z = 0.00$} \\
\cline{2-4} \noalign{\smallskip}
& $b_{\text{eff}}$ & $b_{\text{punct}}$ & $b_{\text{punct}}$ (bc) \\
\midrule
$2\cdot10^{12}$     & $1.122 \pm 0.006$ & $1.383 \pm 0.006$  & $1.383 \pm 0.006$ \\
$2.5\cdot10^{12}$   & $1.140 \pm 0.009$ & $1.390 \pm 0.005$  & $1.397 \pm 0.004$ \\
$5\cdot10^{12}$     & $1.256 \pm 0.011$ & $1.497 \pm 0.008$  & $1.491 \pm 0.007$ \\
$7.5\cdot10^{12}$   & $1.353 \pm 0.011$ & $1.580 \pm 0.014$  & $1.571 \pm 0.009$ \\
$10^{13}$     & $1.429 \pm 0.012$ & $1.641 \pm 0.013$  & $1.644 \pm 0.012$ \\
\end{tabular}

\begin{tabular}{ccccc}
\toprule
\multirow{2}{*}{$M_\textrm{min}$ [${M_\odot /}{h}]$} & \multicolumn{3}{c}{$z = 0.55$} \\
\cline{2-4} \noalign{\smallskip}
& $b_{\text{eff}}$ & $b_{\text{punct}}$ & $b_{\text{punct}}$ (bc) \\
\midrule
$2\cdot10^{12}$     & $1.507 \pm 0.011$ & $1.702 \pm 0.014$  & $1.702 \pm 0.014$ \\
$2.5\cdot10^{12}$   & $1.536 \pm 0.011$ & $1.715 \pm 0.018$  & $1.717 \pm 0.013$ \\
$5\cdot10^{12}$     & $1.730 \pm 0.013$ & $1.915 \pm 0.017$  & $1.893 \pm 0.012$ \\
$7.5\cdot10^{12}$   & $1.872 \pm 0.015$ & $2.062 \pm 0.030$  & $2.032 \pm 0.017$ \\
$10^{13}$     & $2.018 \pm 0.019$ & $2.208 \pm 0.029$  & $2.148 \pm 0.037$ \\
\end{tabular}

\begin{tabular}{ccccc}
\toprule
\multirow{2}{*}{$M_\textrm{min}$ [${M_\odot /}{h}]$} & \multicolumn{3}{c}{$z = 1.00$} \\
\cline{2-4} \noalign{\smallskip}
& $b_{\text{eff}}$ & $b_{\text{punct}}$ & $b_{\text{punct}}$ (bc) \\
\midrule
$2\cdot10^{12}$     & $1.983 \pm 0.017$ & $2.104 \pm 0.017$  & $2.104 \pm 0.017$ \\
$2.5\cdot10^{12}$   & $2.301 \pm 0.017$ & $2.113 \pm 0.017$  & $2.128 \pm 0.036$ \\
$5\cdot10^{12}$     & $2.321 \pm 0.021$ & $2.405 \pm 0.018$  & $2.420 \pm 0.031$ \\
$7.5\cdot10^{12}$   & $2.573 \pm 0.027$ & $2.745 \pm 0.072$  & $2.620 \pm 0.041$ \\
$10^{13}$     & $2.756 \pm 0.031$ & $2.881 \pm 0.028$  & $2.816 \pm 0.033$ \\
\hline
\bottomrule
\end{tabular}

\end{table}

\begin{figure*}
\plotfull{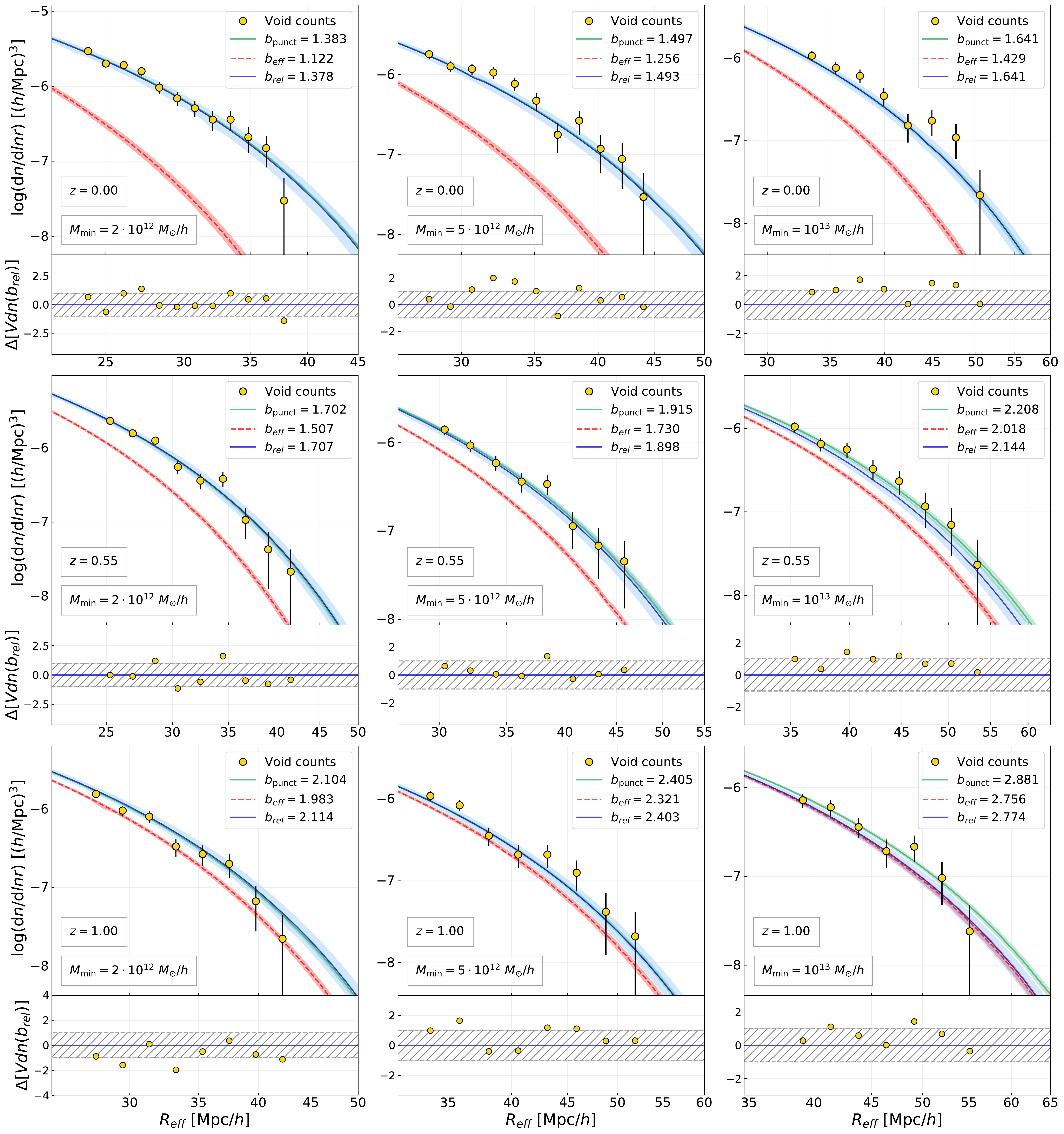}
    \caption{The measured size function of the voids (yellow dots) identified in the DM halo catalogues with $M_\textrm{min} = 2 \cdot 10^{12} \ {M_\odot /}{h}$, $5 \cdot 10^{12} \ {M_\odot /}{h}$ and $10^{13} \ {M_\odot /}{h}$ (rows from \textit{top} to \textit{bottom}), at redshifts $z=0$, $z=0.55$, $z=1$ (columns from \textit{left} to \textit{right}). Voids with $R_{\text{eff}}<2.5$ times the mean inter-particle separation are rejected from the analysis. \textit{Upper sub-panels}: the blue dashed lines represent the void size function obtained by rescaling the Vdn model with $b_{\text{rel}}$, that is the value of the bias computed from the relation shown in Fig. \ref{fig:bias_z}. The green solid lines show the model rescaled with the value of $b_{\text{punct}}$. The red dashed lines represent the model rescaled with the effective bias, $b_{\text{eff}}$. In all cases, the shaded areas indicate the variation of the model obtained applying $1 \sigma$ errors on the value of the tracer bias. \textit{Lower sub-panels}: the residuals of the void counts, computed as the ratio between the difference $\textit{data}-\textit{model}$ and the $\textit{data errors}$, where the data are the measured void size function and the model is given by the re-parameterisation of the Vdn model with $b_{\text{rel}}$.
    The grey dashed areas indicate the regions in which the discrepancy between the data and the model is within the data errors.
    }
    \label{fig:size_functions}
\end{figure*}

However, the $b_{\text{punct}}$ values estimated in the higher bias halo catalogues tend to systematically depart from the fit, at all redshifts. The reason of this slight deviation is related to the method used to find the void centres. In fact, if the detected voids are traced by too few tracers, the {\small VIDE} method might not be sufficiently accurate to localise their centres. Computing the spherically-averaged density contrast starting from a point that is not a local minimum of the density field causes systematic errors in the bias measurements. This is a natural consequence of the cleaning procedure: when rescaling the void radii, the selected threshold might be reached at smaller radii if overdense regions are included in the measurement, due to a bad centering. 
This is an issue especially for catalogues with a high mass selection. 

As a possible strategy to alleviate the problem, we repeat our bias measurements using in all cases the centre positions of the voids detected in the catalogues with the lowest mass-cut. We will refer to this method as our \textit{best-centering} technique, and we will call $b_{\text{punct}} \text{ (bc)}$ the corresponding bias. 
As shown in Fig. \ref{fig:bias_z}, these bias values (shown as coloured circles) are in better agreement with a linear model. Therefore we use them to calibrate the relation between the bias measured on large scales and the one computed inside cosmic voids, obtaining the following equation:
\begin{equation}
  \label{eq:bias_relation}
  b_{\text{punct}} \text{ (bc)} = b_{\text{eff}} \cdot (0.854 \pm 0.007) + (0.420 \pm 0.010) \, .
\end{equation}
This relation can be used to estimate the bias of the tracers inside voids from the effective bias of the whole tracer population. Hereafter, the bias obtained using Eq. \eqref{eq:bias_relation} will be called ${f(b_{\text{eff}})\equiv b_{\text{rel}}}$. All the different bias values are reported in Table \ref{tab:bias}.

It is important to notice that the best-centering technique is not employable with real mocks, since in that case it is not possible to use more numerous tracers to improve the centre of a void. Nevertheless, in our work we choose to rely on this technique to obtain a better calibration of the relation between $b_{\text{punct}}$ and $b_{\text{eff}}$. Indeed, it is convenient to calibrate the latter with $b_\text{punct}$(bc) to minimise the deviation of the data associated to the catalogues with higher mass selections from the linear fit. Using the best-centering technique to alleviate the problem of the sparsity of the tracers, we are able to extend our pipeline also to catalogues with lower spatial resolution.


\begin{figure*}
\plotthree{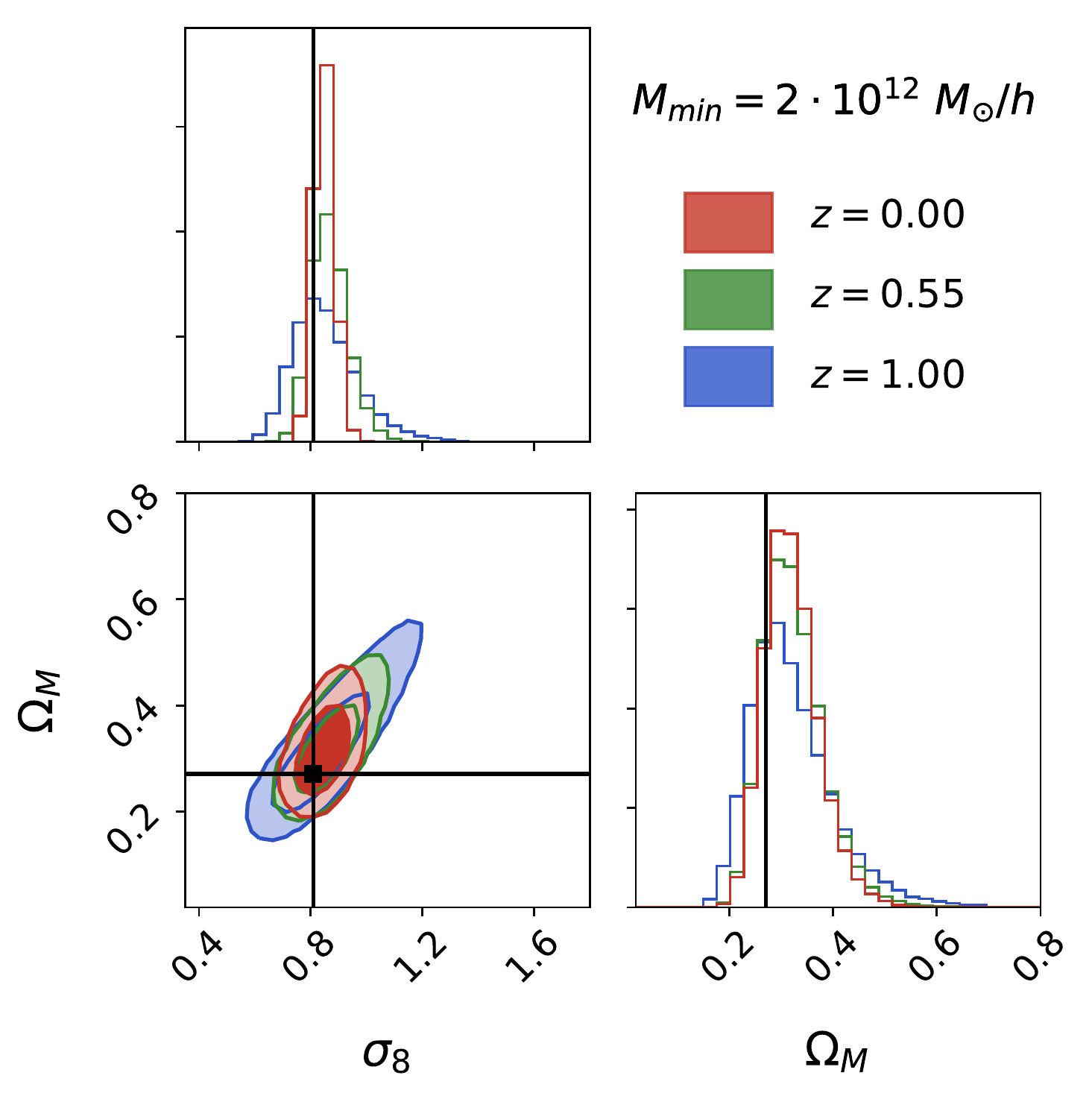}{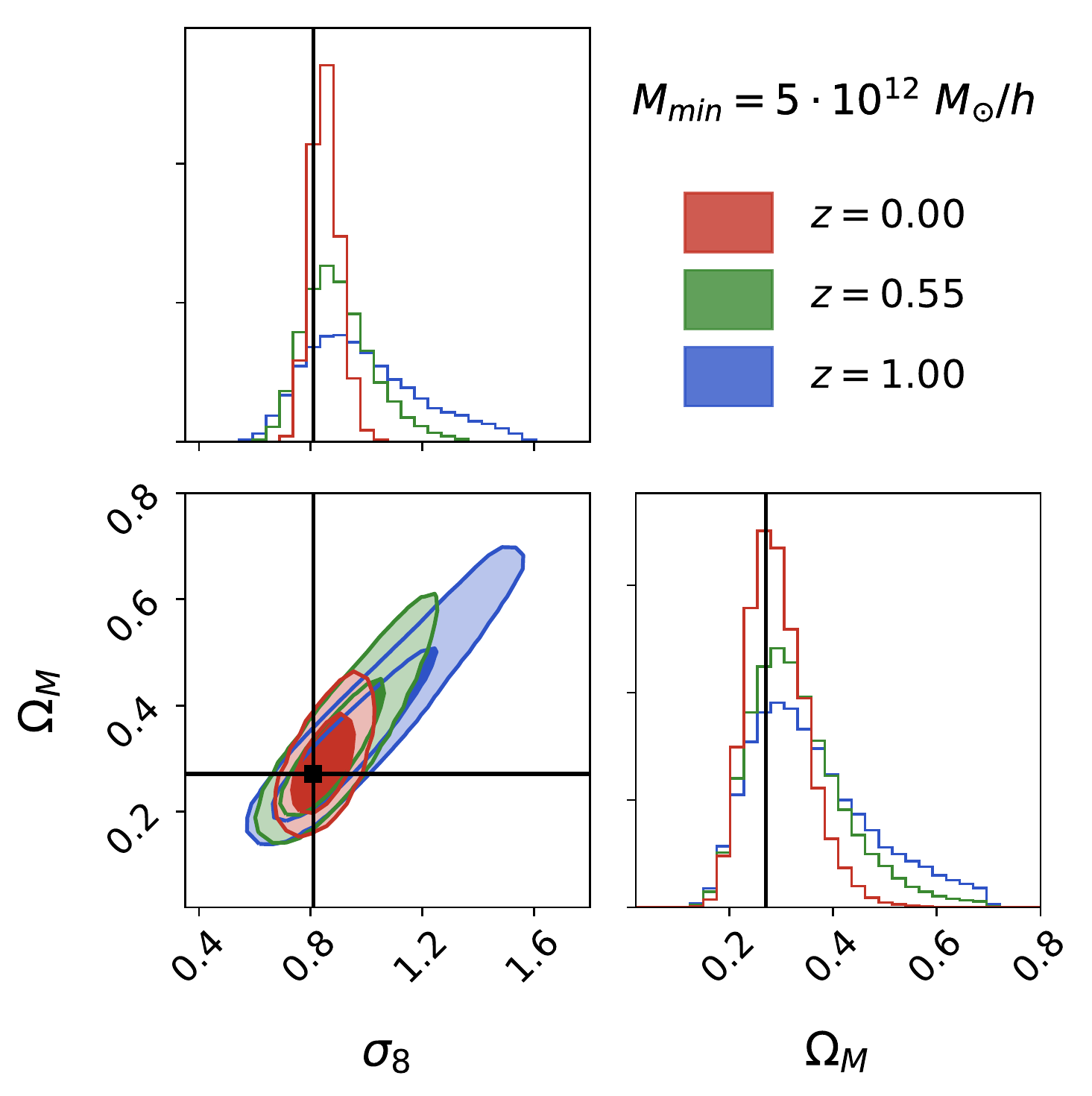}{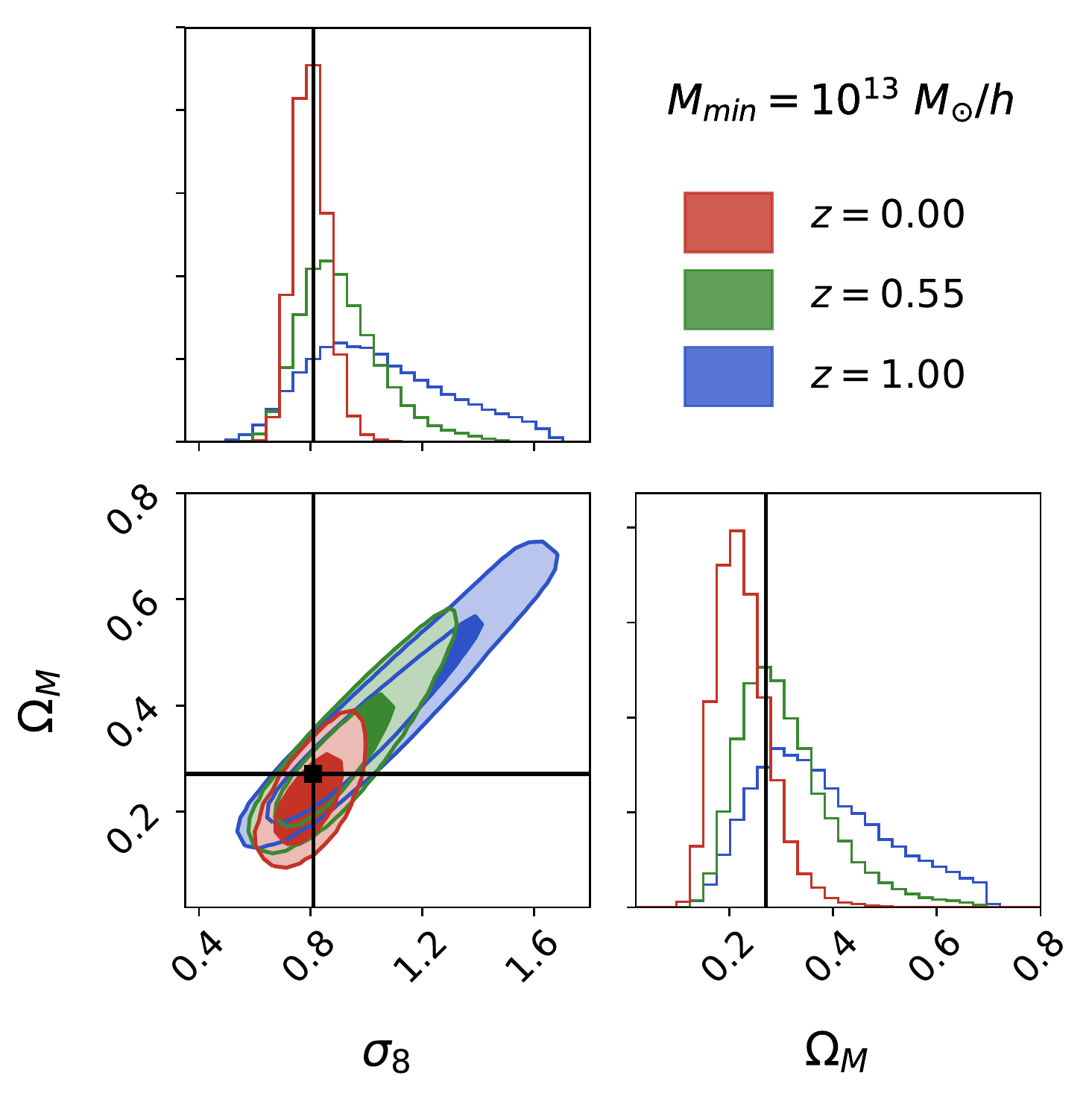}
    \caption{$68\%$ and $95\%$ contours in the $\sigma_8$ - $\Omega_{\rm M}$ plane, for the halo catalogues with $M_\textrm{min} = 2 \cdot 10^{12} \ {M_\odot /}{h}$ (\textit{left}), $5 \cdot 10^{12} \ {M_\odot /}{h}$ (\textit{centre}), and $10^{13} \ {M_\odot /}{h}$ (\textit{right}). The colour of ellipses corresponds to different redshifts: red for $z=0$, green for $z=0.55$ and blue for $z=1$. The prior distributions are uniform for $\sigma_8$ and $\Omega_{\rm M}$, and Gaussian for $b_{\text{eff}}$, $A_{\text{rel}}$ and $B_{\text{rel}}$. The histograms (top and bottom right panels) show the posterior distributions of $\sigma_8$ and $\Omega_{\rm M}$, respectively. The black lines represent the true WMAP7 values ($\sigma_8=0.809$ and $\Omega_{\rm M}=0.2711$).}
    \label{fig:contorni}
\end{figure*}

\subsection{The void size function}
\label{subsec:data-theory}
Here we measure the void size function of our cleaned catalogues and compare it with the theoretical predictions given by the re-parameterised Vdn model. We reject the voids that are too close to the boundaries of the simulation box, as their radii cannot be accurately rescaled by our cleaning algorithm, and we correct consequently the effective volume of the box. The theoretical size function is modelled taking into account the effect of the bias of DM haloes inside voids, as described in Section \ref{subsec:SF_rescaled}. 

Figure \ref{fig:size_functions} displays our results. 
The new re-parameterised void size function model accurately describes all our measurements, in the full range of redshift and mass (thus bias) selections. This represents the main outcome of our analysis. We show both the size function models obtained by rescaling with $b_{\text{punct}}$ and $b_{\text{rel}}$, that appear fully consistent, especially at low redshift and bias values.
The uncertainty in the identification of void centres in low density tracer catalogues causes the slight discrepancies that can be seen at high redshifts and biases, which in any case appear not statistically significant. For comparison, we also show the model obtained by rescaling the Vdn model with the effective bias of the full DM halo population, $b_{\text{eff}}$. As it is clearly evident in the Figure, this case under-predicts systematically the measured size function at all redshifts and biases.


The final goal of this paper is to investigate the cosmological constraints that can be derived from the void size function at different redshifts. To mimic real data analyses, we suppose to have access only to the tracer density field. With no information about the underlying total matter distribution, we have to rely on the relation found in Section \ref{subsec:relation_bias}. We first estimate the effective bias of the sample, $b_{\text{eff}}$, and we consider the coefficients shown in Eq. \eqref{eq:bias_relation}, $A_{\text{rel}}$ and $B_{\text{rel}}$, that are the offset and the slope of the calibrated relation, respectively. These coefficients are necessary to convert $b_{\text{eff}}$ into $b_{\text{rel}}$, which in turn is required to re-parameterise the Vdn model, as shown in Fig \ref{fig:size_functions}. Then, we perform a Bayesian statistical MCMC analysis of the measured void size function by sampling the posterior distribution of the parameters $\sigma_8$ and $\Omega_{\rm M}$. We assume uniform prior distributions for $\sigma_8$ and $\Omega_{\rm M}$, and we leave as free parameters also $b_{\text{eff}}$, $A_{\text{rel}}$ and $B_{\text{rel}}$, assuming in this case Gaussian prior distributions centered at the estimated values of these parameters, with standard deviations equal to their relative $1 \sigma$ uncertainties.
The results for our simulated catalogues with three different mass-cuts and redshifts are reported in Fig. \ref{fig:contorni}. The true values of the cosmological parameters are within the $68\%$ levels in all cases.
In Appendix \ref{Appendix_C} we investigate the systematics in the cosmological constraints possibly caused by the uncertainites in the estimation of the tracer bias, while in Appendix \ref{Appendix_D} we show the outcome of combining the posterior distributions at different redshifts to achieve tighter constraints on the cosmological parameters.

\begin{figure*}
\plotfull{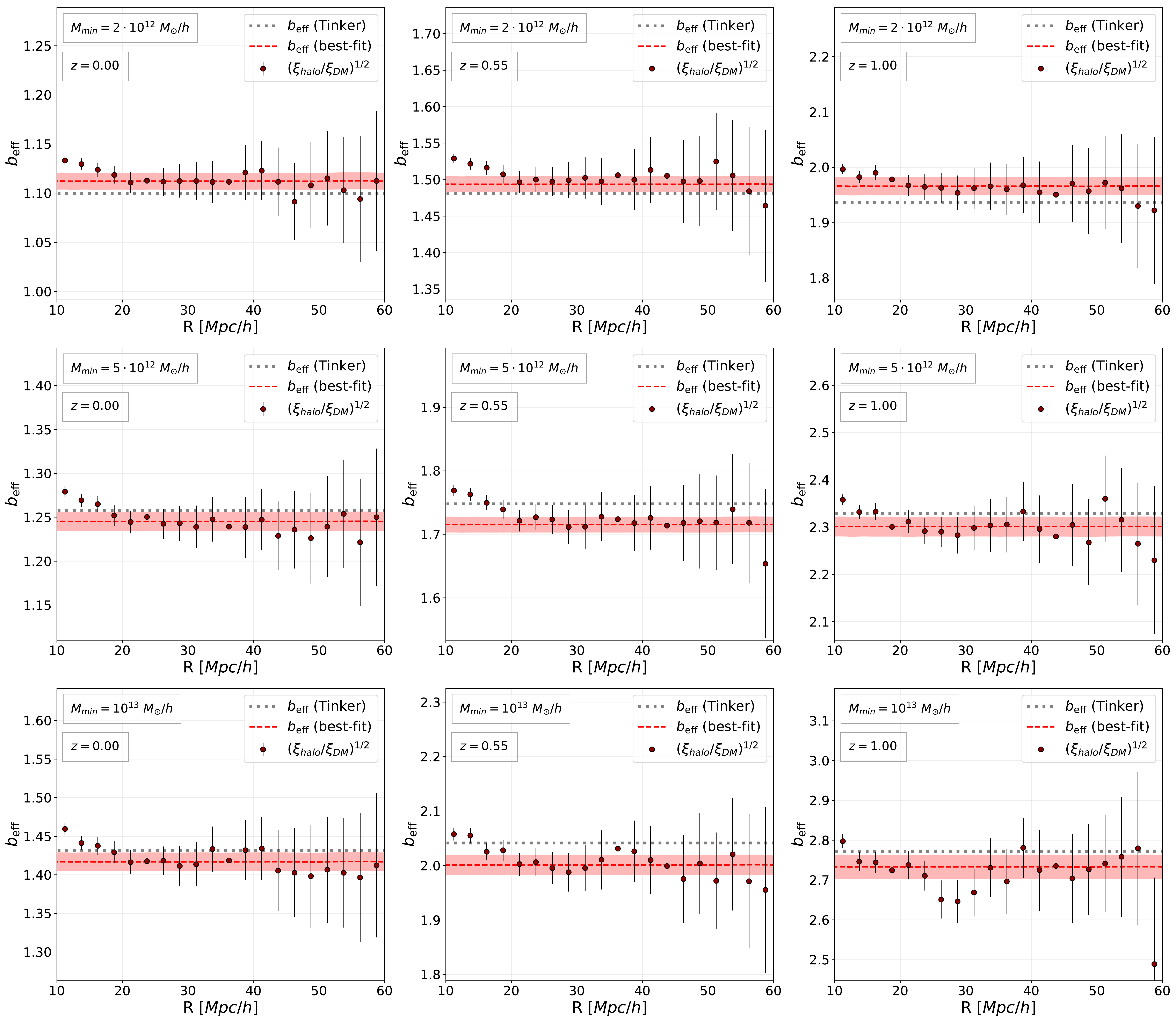}
    \caption{The halo bias for the catalogues with $M_{\textrm{min}} = 2 \cdot 10^{12} \ {M_\odot /}{h}$, $5 \cdot 10^{12} \ {M_\odot /}{h}$ and $10^{13} \ {M_\odot /}{h}$ (rows from \textit{top} to \textit{bottom}), at redshifts $z=0$, $z=0.55$, $z=1$ (columns from \textit{left} to \textit{right}). The black points represent the square root of the ratio between the auto-correlation function of the haloes and the DM particles (Eq. \eqref{auto-corr_ratio}). The error bars are the diagonal elements of the covariance matrix estimated with Bootstrap. The red shaded areas show the $1 \sigma$ uncertainties on the best-fit bias values estimated with the MCMC modelling, fitting in the range of radii of $[20$-$40]$ $\mbox{Mpc}/h$. The dashed grey lines show the theoretical predictions given by the \citet{Tinker_bias} model.}
    \label{fig:bias_modelling}
\end{figure*}


\section{Conclusions}
\label{sec:conclusions}
In this work, we have extended the prescriptions developed by \citet{Roncarini2019} to model cosmic void abundances. In particular, we have focused on the relation between the size-abundance distribution of voids and the bias of the tracers used to detect them.
We have also investigated the cosmological constraints that can be derived from void number counts at different redshifts.  

The main steps and results of our work can be summarised as follows.
\begin{itemize}
    \item We have run the finding and cleaning algorithms on simulated DM halo catalogues, selected with different mass-cuts to probe different tracer biases and redshifts.
    \item We have compared the tracer bias measured on large scales to the one measured inside cosmic voids, finding a tight relation (see Eq. \ref{eq:bias_relation}) between the two.
    \item Then we have provided a new parameterisation of the Vdn model as a function of the large-scale tracer bias. We have calibrated the model on our simulated catalogues at different redshifts and biases.  
    \item Finally, we have explored the constraining power of the void size function. Specifically, we have performed a Bayesian statistical inference analysis, fitting the measured size function with the new calibrated model, obtaining constraints on $\sigma_8$ and $\Omega_{\rm M}$.
\end{itemize}
In this paper we have investigated one of the possible cosmological applications of cosmic void statistics, that is void abundances.
The bias relation calibrated in Section \ref{subsec:relation_bias} allows to construct the likelihood for the statistical inference analysis as a function of the large-scale effective bias of the sample. This work lays the foundations for the cosmological exploitation of the void size function, when the voids are identified in the distribution of biased tracers, such as in real data catalogues.

We note that another interesting application of our method would be to combine results using tracers with significantly different bias, e.g. combining optical surveys to HI surveys, where the bias can be negative \citep{negative_bias1, negative_bias2}. This test is beyond the scope of the paper and we leave it as a future development of our work.

Finally it is important to notice that the relation between $b_{\text{punct}}$ and $b_{\text{eff}}$ provided in Eq. \eqref{eq:bias_relation} is valid in the $\Lambda$CDM framework only. Extending our method to other cosmological scenarios (e.g. for constraining modified DE models) requires to calibrate the relation using appropriate N-body simulations.


\section*{Acknowledgements}
We acknowledge the anonymous referee for the useful comments, which significantly contributed to the improvement of the quality of the manuscript.
TR is grateful for the support of his supervisors, Andrea Lapi and Matteo Viel.
FM and LM acknowledge the grants ASI n.I/023/12/0, ASI-INAF n.
2018-23-HH.0 and PRIN MIUR 2015 ``Cosmology and Fundamental Physics: illuminating the Dark Universe with Euclid''.


\appendix

\section{Measuring the linear bias}
\label{Appendix_A}

\begin{figure*}
\plotfull{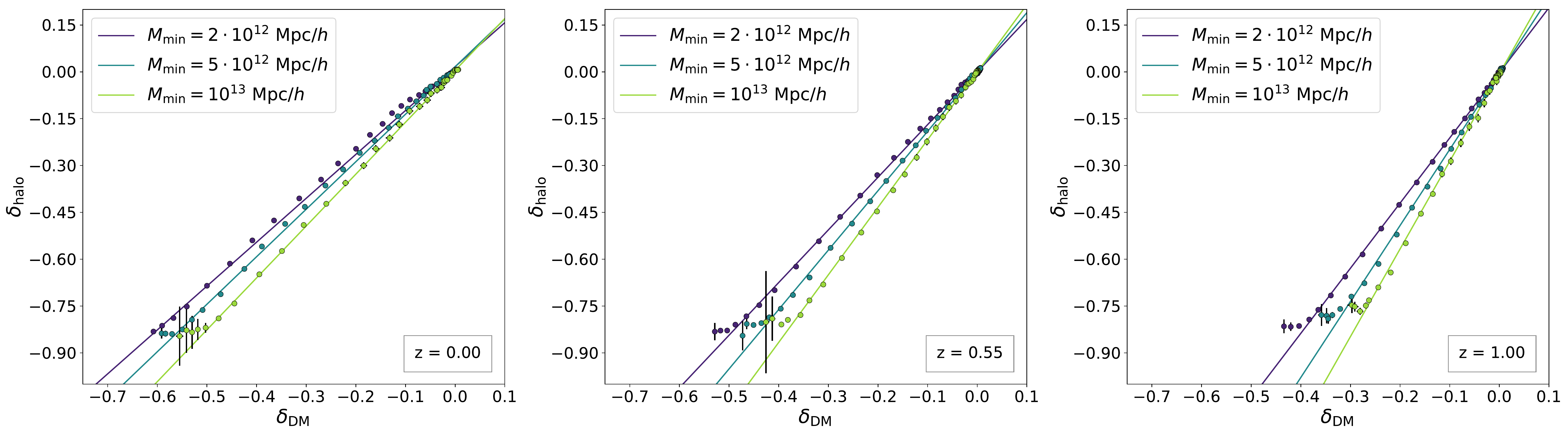}
    \caption{The ratio of the stacked density profiles shown in Fig. \ref{fig:density_profiles}, that is $\delta_{v,\, \text{DM}}^{NL}$ as a function $\delta_{v,\, \text{tr}}^{NL}$, at redshifts $z=0$ (\textit{left}), $z=0.55$ (\textit{centre}), $z=1$ (\textit{right}). Different colours correspond to the halo catalogues with $M_{\textrm{min}} = 2 \cdot 10^{12} \ {M_\odot /}{h}$ (in violet), $M_{\textrm{min}} = 5 \cdot 10^{12} \ {M_\odot /}{h}$ (in blue) and $M_{\textrm{min}} = 10^{13} \ {M_\odot /}{h}$ (in green). The black error bars represent $1 \sigma$ uncertainties. As expected, the slope of the fit becomes steeper with higher mass-cuts (thus, the value of bias inside voids). }
    \label{fig:b_slope}
\end{figure*}

\begin{table}
\centering
\caption{Table of the values $b_{\text{slope}}$ and $c_{\text{off}}$ obtained by the linear fitting of $\delta_\text{DM}$ and $\delta_\text{halo}$, as shown in Fig. \ref{fig:b_slope}. Here are presented the results for all the halo catalogues (with five mass cuts) at redshifts $z=0$, $z=0.55$ and $z=1$. We report also the values of $\tilde{b}_{\text{punct}}$ computed with Eq. \eqref{eq:b_punct_tilde}.}
\begin{tabular}{ccccc}
\toprule
\multirow{2}{*}{$M_\textrm{min}$ [${M_\odot /}{h}]$} & \multicolumn{3}{c}{$z = 0.00$} \\
\cline{2-4} \noalign{\smallskip} & $b_{\text{slope}}$ & $c_{\text{off}}$& $\tilde{b}_{\text{punct}}$\\
\midrule
$2\cdot10^{12}$     & $1.405 \pm 0.009$ & $0.017 \pm 0.004$  & $1.373 \pm 0.011$\\
$2.5\cdot10^{12}$   & $1.415 \pm 0.008$ & $0.015 \pm 0.003$  & $1.386 \pm 0.010$\\
$5\cdot10^{12}$     & $1.521 \pm 0.009$ & $0.016 \pm 0.003$  & $1.486 \pm 0.011$\\
$7.5\cdot10^{12}$   & $1.562 \pm 0.007$ & $0.002 \pm 0.002$  & $1.561 \pm 0.009$\\
$10^{13}$           & $1.661 \pm 0.006$ & $0.005 \pm 0.002$  & $1.650 \pm 0.008$\\
\end{tabular}

\begin{tabular}{ccccc}
\toprule
\multirow{2}{*}{$M_\textrm{min}$ [${M_\odot /}{h}]$} & \multicolumn{3}{c}{$z = 0.55$} \\
\cline{2-4} \noalign{\smallskip} & $b_{\text{slope}}$ & $c_{\text{off}}$& $\tilde{b}_{\text{punct}}$\\
\midrule
$2\cdot10^{12}$     & $1.685 \pm 0.009$ & $-0.001 \pm 0.003$    & $1.687 \pm 0.011$\\
$2.5\cdot10^{12}$   & $1.789 \pm 0.010$ & $0.004 \pm 0.003$     & $1.719 \pm 0.012$\\
$5\cdot10^{12}$     & $1.901 \pm 0.012$ & $-0.001 \pm 0.003$    & $1.903 \pm 0.014$\\
$7.5\cdot10^{12}$   & $2.032 \pm 0.019$ & $-0.002 \pm 0.004$    & $2.037 \pm 0.023$\\
$10^{13}$           & $2.163 \pm 0.013$ & $-0.001 \pm 0.003$    & $2.167 \pm 0.016$\\
\end{tabular}

\begin{tabular}{ccccc}
\toprule
\multirow{2}{*}{$M_\textrm{min}$ [${M_\odot /}{h}]$} & \multicolumn{3}{c}{$z = 1.00$} \\
\cline{2-4} \noalign{\smallskip} & $b_{\text{slope}}$ & $c_{\text{off}}$& $\tilde{b}_{\text{punct}}$\\
\midrule
$2\cdot10^{12}$     & $2.086 \pm 0.010$ & $-0.003 \pm 0.002$  & $2.095 \pm 0.011$\\
$2.5\cdot10^{12}$   & $2.129 \pm 0.011$ & $-0.005 \pm 0.002$  & $2.145 \pm 0.012$\\
$5\cdot10^{12}$     & $2.433 \pm 0.019$ & $-0.005 \pm 0.003$  & $2.452 \pm 0.021$\\
$7.5\cdot10^{12}$   & $2.690 \pm 0.016$ & $-0.005 \pm 0.002$  & $2.711 \pm 0.017$\\
$10^{13}$           & $2.806 \pm 0.025$ & $-0.006 \pm 0.003$  & $2.830 \pm 0.028$\\

\hline
\bottomrule
\end{tabular}
\label{tab:b_slope}
\end{table}

In this Appendix, we describe the methods employed in this work to estimate the large-scale effective linear bias of the tracers used to identify the voids. We followed the same prescriptions as in \citet{Marulli2013, Marulli2018}, exploiting the 2PCF of the DM haloes of our simulated catalogues, and performing a Bayesian statistical analysis to infer the effective bias, $b_{\text{eff}}$.

The angle-averaged 2PCF $\hat{\xi}(r)$ is computed using the \citet{Landy_Szalay1993} estimator:
\begin{equation} \label{2PCF_estimator}
\hat{\xi}(r)=\frac{N_{RR}}{N_{OO}}\frac{OO(r)}{RR(r)}-2\frac{N_{RR}}{N_{RR}}\frac{OR(r)}{RR(r)}+1 \textrm{ ,}
\end{equation}
where $OO(r)$, $RR(r)$ and $OR(r)$ are the binned numbers of object-object, random-random, and object-random pairs with distance $r\pm\Delta r$, while $N_{OO}=N_O(N_O-1)/2$, $N_{RR}=N_R(N_R-1)/2$ and $N_{RR}=N_O N_R$ are the total numbers of object-object, random-random, and object-random pairs in the sample, respectively, and $N_O$ and $N_R$ are the total number of objects and random objects, respectively. The \citet{Landy_Szalay1993} estimator provides an unbiased estimate of the 2PCF in the limit $N_R \rightarrow \infty$, with minimum variance.

\begin{figure*}
\plotthree{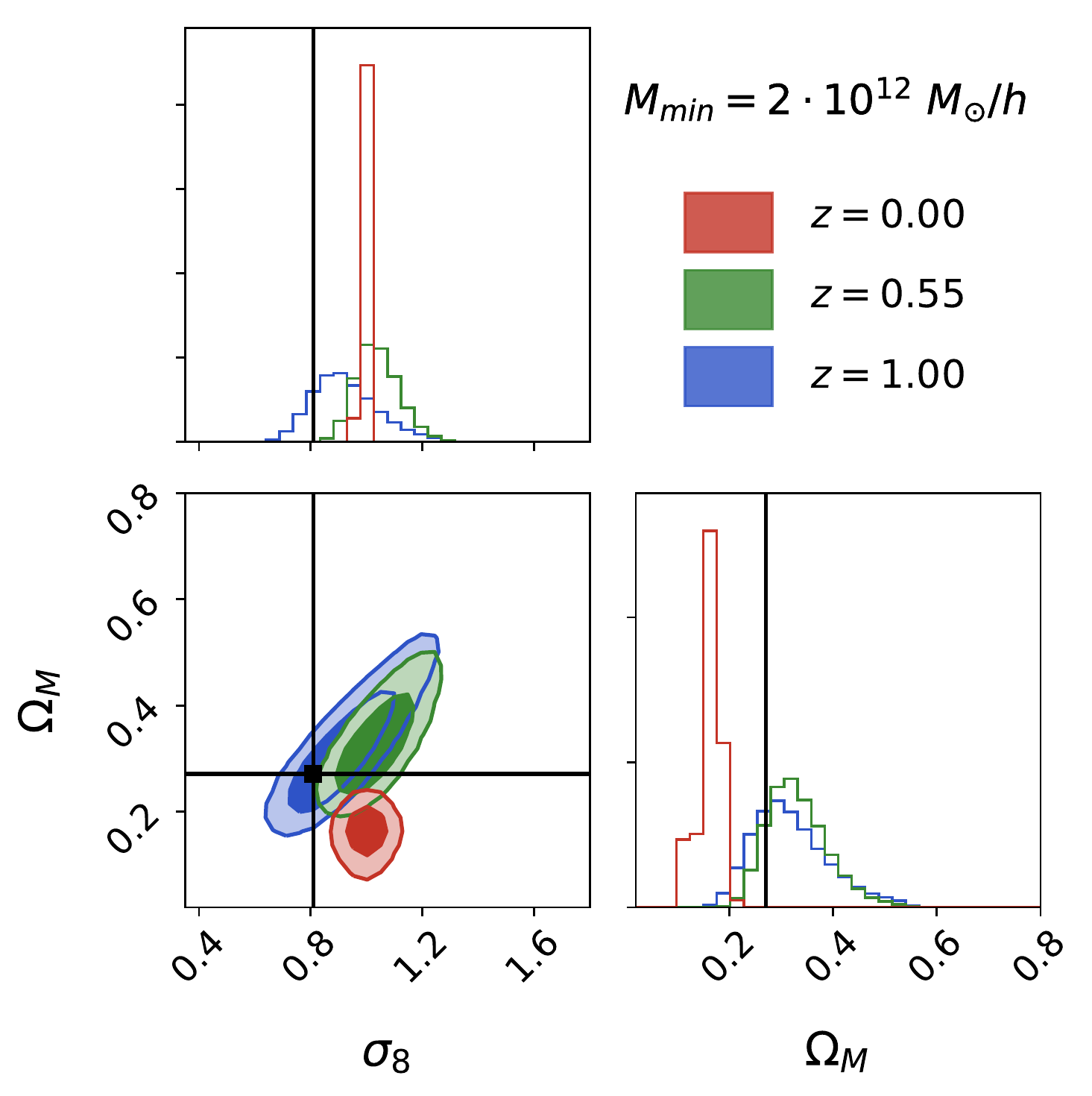}{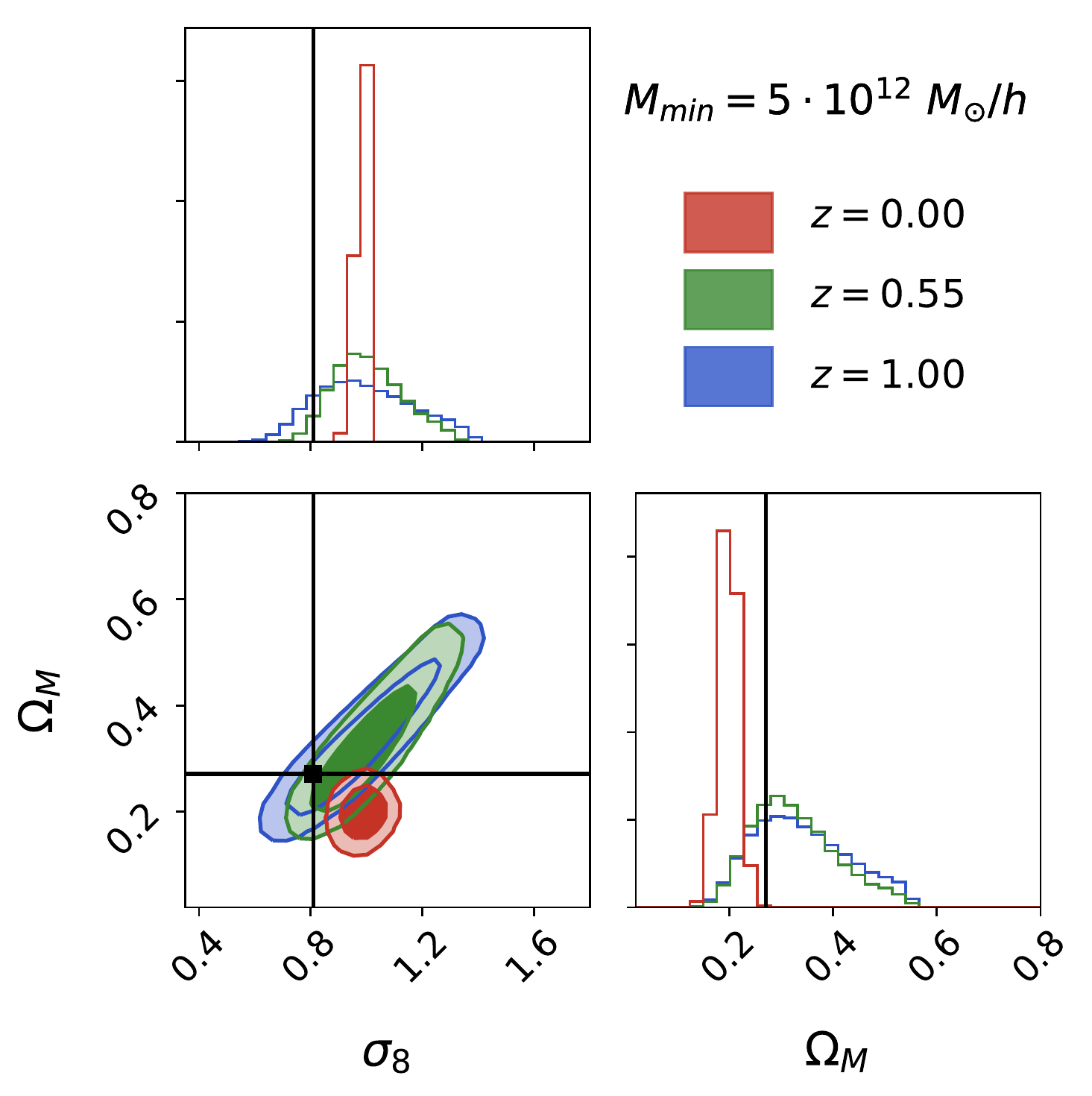}{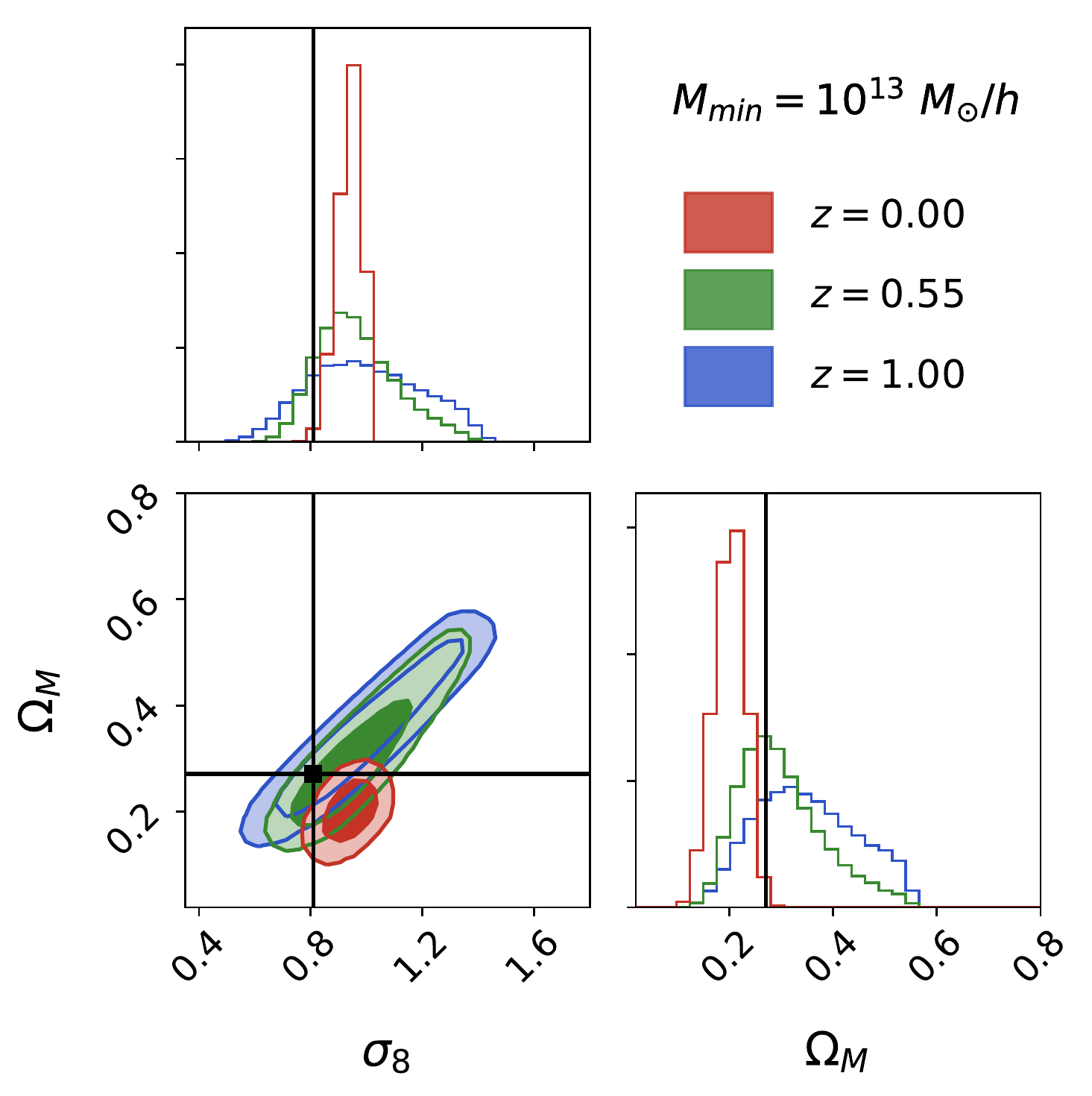}
    \caption{$68\%$ and $95\%$ contours in the $\sigma_8$ - $\Omega_{\rm M}$ plane, for the halo catalogues with $M_\textrm{min} = 2 \cdot 10^{12} \ {M_\odot /}{h}$ (\textit{left}), $5 \cdot 10^{12} \ {M_\odot /}{h}$ (\textit{centre}), and $10^{13} \ {M_\odot /}{h}$ (\textit{right}), obtained by re-parameterising the Vdn model directly with $b_{\text{eff}}$, thus without converting this value by means of the Eq. \eqref{eq:bias_relation}. The colour of ellipses corresponds to different redshifts: red for $z=0$, green for $z=0.55$ and blue for $z=1$. The prior distributions are uniform for $\sigma_8$ and $\Omega_{\rm M}$, and Gaussian for $b_{\text{eff}}$. The histograms (top and bottom right panels) show the marginalised posterior distributions of $\sigma_8$ and $\Omega_{\rm M}$, respectively. The black lines represent the true WMAP7 values ($\sigma_8=0.809$ and $\Omega_{\rm M}=0.2711$).}
    \label{fig:contorni_beff}
\end{figure*}

We computed the covariance matrix $C_{i,j}$, which measures the variance and correlation between the different bins of the 2PCF, with the Bootstrap method, dividing the original catalogues in 125 sub-catalogues, and constructing 100 realisations by resampling from the sub-catalogues, with replacement. We constructed the random catalogue by extracting the object coordinates randomly, preserving the same three-dimensional coverage and the same geometry of the initial catalogue. In particular, we build the random catalogue to be four times larger than the DM halo sample, since this proportion allows to have sufficiently small Poissonian errors in the DR counts, compared to the errors in DD. We also performed tests with different sizes of the random catalogue, finding consistent results.

The covariance matrix is defined as follows:
\begin{equation}
C_{i,j}=\mathcal{F} \sum_{k=1}^{N_R}(\xi_i^k-\overline{\xi}_i)(\xi_j^k-\overline{\xi}_j) \textrm{ ,}
\end{equation}
where the subscripts $i$ and $j$ run over the 2PCF bins, while $k$ refers to the 2PCF of the $k\textrm{-}th$ of $N_R$ catalogue realisations, and $\hat{\xi}$ is the mean 2PCF of the $N_R$ samples. $\mathcal{F}$ is the normalisation factor, which takes into account the fact that the $N_R$ realisations might not be independent \citep[][]{Norberg2019}, and is $\mathcal{F}=1/(N_R-1)$ in the case of the Bootstrap method. 

Finally, we performed a full MCMC analysis of the 2PCF, using a Gaussian likelihood function $\mathcal{L}$, defined as:
\begin{equation}\label{likelihood}
-2\mbox{ln}\mathcal{L}=\sum^N_{i=1}\sum^N_{j=1}(\xi_i^d-\xi_i^m)C_{i,j}^{-1}(\xi_j^d-\xi_j^m) \textrm{ ,}
\end{equation}
where $C^{-1}_{i,j}$ is the inverse of the covariance matrix, $N$ is the number of comoving separation bins at which the 2PCF is estimated,
and the superscripts $d$ and $m$ stand for data and model, respectively.  The 2PCF model, $\xi^m(r)$, is computed as follows:
\begin{equation}
\xi_m(r)= b^2_{\text{eff}}\,\xi_{\text{DM}}(r) \textrm{ ,}
\end{equation}
where $\xi_{\text{DM}}(r)$ is the DM 2PCF, which is estimated by Fourier transforming the power spectrum, $P_{\text{DM}}(k)$, computed with the {\small Code for Anisotropies in the Microwave Background} ({\small CAMB}, \hyperlink{CAMB}{http://camb.info}).
An accurate estimate of the effective bias parameter, $b_{\text{eff}}$, and its uncertainty are assessed by sampling its posterior distribution.

Figure \ref{fig:bias_modelling} shows the results of this analysis. The data points are the square root of ratio between the tracer and matter 2PCFs:
\begin{equation} \label{auto-corr_ratio}
b = \sqrt{\xi_{tr} / \xi_{m}}\, ,
\end{equation}
while the dashed red lines show the best-fit values and uncertainties of $b_{\text{eff}}$, estimated from the median and quartiles of the posterior distribution sampled with the MCMC analysis.

We compared these values to the theoretical effective bias of DM haloes, computed as follows: 
\begin{equation}
\label{b_eff}
b_{\text{eff}}(z) = \frac{\int_{M_{\mathrm{min}}}^{M_{\mathrm{max}}} {\rm d}M\, b(M, z) \Phi(M, z)}{\int_{M_{\mathrm{min}}}^{M_{\mathrm{max}}} {\rm d}M\,\Phi(M, z)} \textrm{ ,}
\end{equation}\\
where $\Phi(M, z)$ is the halo mass function of the catalogue, and $M_{\mathrm{min}}$ and $M_{\mathrm{max}}$ are the lowest and largest masses in the sample, respectively. To compute the linear bias $b(M, z)$, we relied on the theoretical model developed by \citet{Tinker_bias}.


\section{Testing different bias estimates inside voids}
\label{Appendix_B}

\citet{pollinarelative} found a linear relation between the spherically-averaged density profiles of biased tracers inside voids, $\delta_{v,\, \text{tr}}^{NL}$, and the underlying DM, $\delta_{v,\, \text{DM}}^{NL}$. The slope of this relation provides an alternative estimate of the bias of void tracers. We repeated the analysis of \citet{pollinarelative} finding consistent results, as shown in Fig. \ref{fig:b_slope}, that reports this relation measured in three DM halo catalogues at three different redshifts.

Even if the offset of the linear fit is small, it can be taken into account to recover a better estimate of the bias inside voids. In particular, to obtain a value comparable with $b_{\text{punct}}$ (see Eq. \eqref{eq:b_punct}), we can recover the bias at a distance of one effective radius, $R_{\text{eff}}$, from void centres, that is where $\delta_{v,\, \text{tr}}^{NL}=-0.7$ (thus the requested threshold we chose for the cleaning procedure). Specifically, from the fitting of the data we obtained:
\begin{equation}
  \label{eq:b_slope}
   \delta_{v,\, \text{tr}}^{NL} = b_{\text{slope}} \ \delta_{v,\, \text{DM}}^{NL} + c_{\text{off}}\,,
\end{equation}
and we computed Eq. \eqref{eq:b_slope} at $R=R_{\text{eff}}$, imposing $\delta_{v,\, \text{tr}}^{NL}(R_{\text{eff}})=-0.7$. Then, dividing both sides by $\delta_{v,\, \text{tr}}^{NL}(R_{\text{eff}})$, we can derive a bias value equivalent to $b_{\text{punct}}$ as follows:
\begin{equation}
  \label{eq:b_punct_tilde}
   \tilde{b}_{\text{punct}} \equiv \frac{b_{\text{slope}}}
   {1-\frac{c_{\text{off}}}{\delta_{v,\, \text{tr}}^{NL}(R_{\text{eff}})}} \, .
\end{equation}
In Table \ref{tab:b_slope} we report the results obtained for all the analysed halo catalogues. As expected, we found similar values for $\tilde{b}_{\text{punct}}$ and $b_{\text{punct}}$. Nevertheless, the method described here is not particularly accurate for the following reasons (see Fig. \ref{fig:bias_compare} as a reference): 
\begin{itemize}
\item First, using the stacked profiles of voids we cannot distinguish a possible variation of the bias as a function of void radii.
\item Second, the slope of the fit depends on the radial extension of the profiles: the wider the regions are embedded, the more the bias computed will tend to the one of the overdensities. There is not a preferential value for the maximum radius of the profiles.
\item Third, a linear fit to the relation between $\delta_\text{halo}$ and $\delta_{\text{DM}}$ is not accurate enough in the full range of $\delta_{\text{DM}}$.
\end{itemize}
This method can be refined considering different bins of void radii and using different linear fits for each of them, to account for the variation of the bias as a function of $R_{\text{eff}}$.

\begin{figure}
\plotone{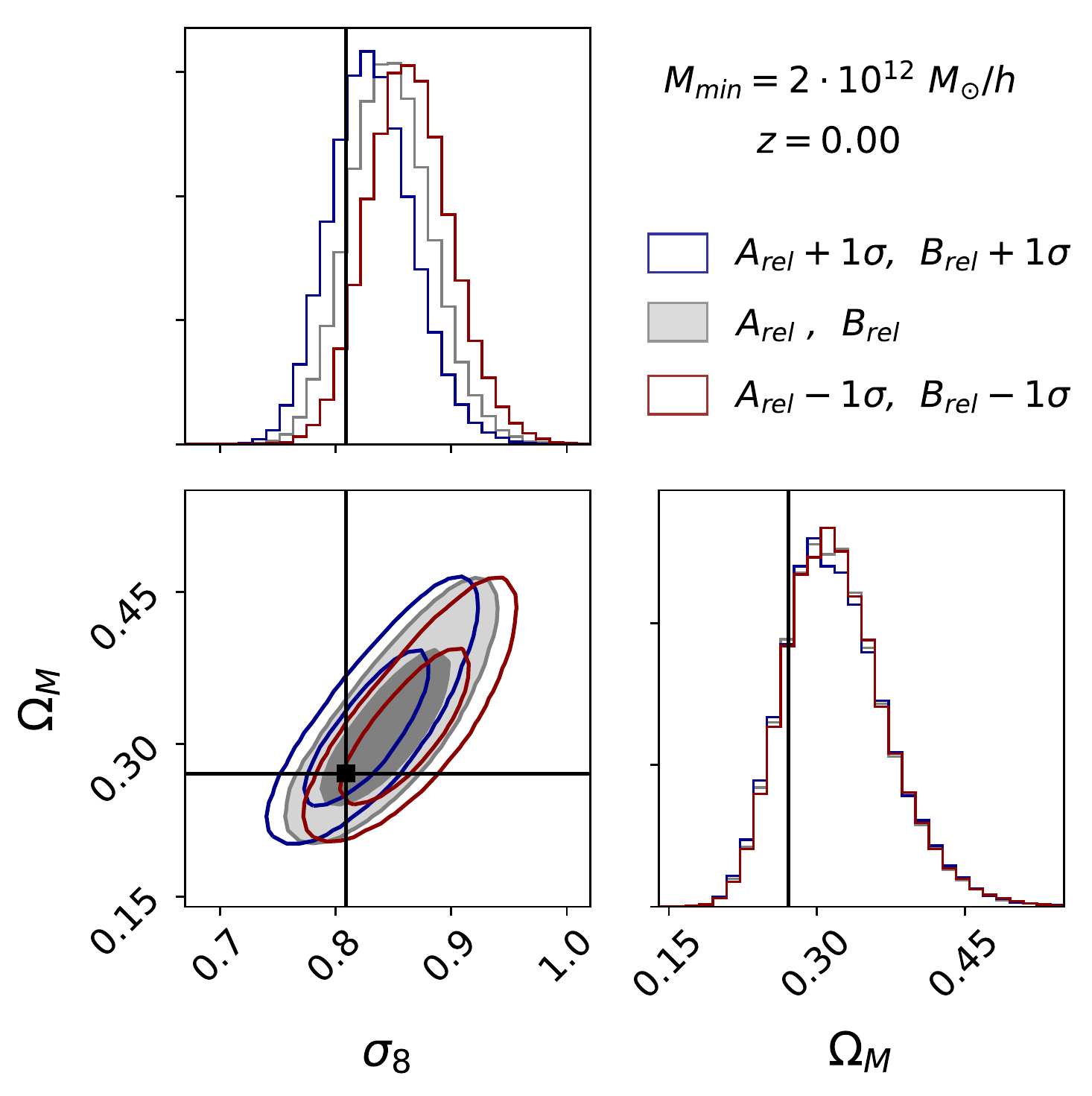}
    \caption{$68\%$ and $95\%$ contours in the $\sigma_8$ - $\Omega_{\rm M}$ plane, for the halo catalogue with $M_\textrm{min} = 2 \cdot 10^{12} \ {M_\odot /}{h}$ at $z=0$. The histograms (top and bottom right panels) show the posterior distributions of $\sigma_8$ and $\Omega_{\rm M}$, respectively. The grey filled contours represent the confidence levels obtained using Eq. \eqref{eq:bias_relation}, while the blue and red contours indicate the results obtained by converting the value of $b_{\text{eff}}$ shifting both the values of $A_{\text{rel}}$ and $B_{\text{rel}}$ by $+1\sigma$ and $-1\sigma$, respectively.} The black lines represent the true WMAP7 values ($\sigma_8=0.809$ and $\Omega_{\rm M}=0.2711$).
    \label{contorni}
\end{figure}

\section{Testing systematics caused by uncertainties on the calibrated bias relation}
\label{Appendix_C}

Here we first test the effect of using the value of $b_{\text{eff}}$ instead of $b_{\text{rel}}$ to recover the cosmological parameters. In particular, we repeat the MCMC analysis of the measured size functions performed in Section \ref{subsec:data-theory} employing a wrong theoretical model, that is a Vdn model re-scaled with the linear bias inferred from the tracer large-scale 2PCF. As demonstrated in Fig. \ref{fig:size_functions}, the model obtained using the tracers bias $b_{\text{eff}}$ cannot fit properly the measured void abundances, unless this value is previously converted in $b_{\text{rel}}$ by means of the relation in Eq. \eqref{eq:bias_relation}. As shown in Fig. \ref{fig:contorni_beff}, the contour levels achieved with the effective bias are on average smaller with respect to the ones presented in Fig. \ref{fig:contorni}. In fact, the uncertainties associated to the theoretical model re-parameterised directly with $b_{\text{eff}}$ are smaller, since the errors of $A_{\text{rel}}$ and $B_{\text{rel}}$ are not included in the model. As expected, the contour levels obtained using the wrong bias value tend to shift from the real values of $\sigma_8$ and $\Omega_{\rm M}$, especially for low redshifts and mass-cuts. Indeed, in these cases the values of $b_{\text{eff}}$ and $b_{\text{rel}}$ are significantly different from each other, whereas at higher redshifts and mass-cuts they tend to be more similar, as showed in Fig. \ref{fig:bias_R}.

We secondly assess the systematic errors on the cosmological constraints caused by uncertainties in the estimation of the coefficients of the conversion relation, calibrated in Section \ref{subsec:relation_bias}. This is particularly useful in the perspective of a future application on real surveys. To propagate a possible systematic error on the Eq. \eqref{eq:bias_relation} to the final cosmological constraints, we repeated the MCMC analysis described in Section \ref{subsec:data-theory} assuming different values for the coefficient $A_{\text{rel}}$ and $B_{\text{rel}}$. In particular, to test the cases with the major discrepancy from the calibrated relation, we increased or decreased both the parameters by $1\sigma$, where $1\sigma$ is the uncertainty derived by the weighted fit of the data in Fig. \ref{fig:bias_z}. Specifically, we set $A'_{\text{rel}}=0.420+0.010$ and $B'_{\text{rel}}=0.854+0.007$ in the first case, whereas $A''_{\text{rel}}=0.420-0.010$ and $B''_{\text{rel}}=0.854-0.007$ in the second case.
In Fig. \ref{contorni} we report the results for the catalogue with $M_\textrm{min} = 2 \cdot 10^{12} \ {M_\odot /}{h}$ at $z=0$.
As shown in this Figure, the real values of $\sigma_8$ and $\Omega_{\rm M}$ are within the $68\%$ confidence levels obtained in both cases.
Moreover, the posterior distribution of $\Omega_{\rm M}$ is almost unchanged, while $\sigma_8$ results shifted towards greater values using a conversion relation with $A''_{\text{rel}}$ and $B''_{\text{rel}}$ and towards lower values for the case with  $A'_{\text{rel}}$ and $B'_{\text{rel}}$. We obtained the same results also for the catalogue with higher redshift and mass selections. The larger is the tracer bias, the larger is the discrepancy of the modified relation from the one calibrated in Eq. \eqref{eq:bias_relation}. Indeed, shifting both the values of $A_{\text{rel}}$ and $B_{\text{rel}}$ by $+1\sigma$ and $-1\sigma$, the resulting linear equations tend to move even further away from the calibrated relation with $b_\text{eff}$. This causes a systematic error that has more impact on the theoretical size functions associated to the catalogues with higher $b_\text{eff}$. 
Nevertheless, we verified that even in these cases the constraints are still consistent with the real values of $\sigma_8$ and $\Omega_{\rm M}$.
We can conclude that, even with a systematic error of $\pm 1\sigma$ on the values of the coefficients in the calibrated relation, the void size function still provides reliable cosmological constraints.

\section{Combining samples at different redsfhits}
\label{Appendix_D}

\begin{table}
\centering
\caption{Mean and standard deviation of the posterior distributions for the parameters $\sigma_8$ and $\Omega_{\rm M}$, computed from the Bayesian statistical analysis of the measured void size functions for the DM halo catalogues with $M_{\textrm{min}} = 2 \cdot 10^{12} \ {M_\odot /}{h}$, $5 \cdot 10^{12} \ {M_\odot /}{h}$ and $10^{13} \ {M_\odot /}{h}$ at $z=0$, $z=0.55$ and $z=1$. The last line of each table reports the results obtained by combining the posterior distributions at the three different redshifts.}
\begin{tabularx}{0.8\columnwidth}{Xcc|cc}

\hline
\toprule
\rowcolor{Gray}
\multicolumn{5}{c}{$M_\textrm{min} = 2\cdot10^{12} {M_\odot /}{h}$}  \\
\hline 
\noalign{\smallskip}
{} & \multicolumn{2}{c}{$\sigma_8$} & \multicolumn{2}{c}{$\Omega_{\rm M}$}\\
\noalign{\smallskip}
\cline{2-3}\cline{4-5}
{} & \textrm{Mean} & \textrm{St. dev.} & \textrm{Mean} & \textrm{St. dev.} \\
\cline{2-5}
\noalign{\smallskip}
$z=0.00$                 & $0.848$ & $0.036$ & $0.321$ & $0.052$  \\
$z=0.55$                 & $0.868$ & $0.068$ & $0.325$ & $0.059$  \\
$z=1.00$                 & $0.856$ & $0.118$ & $0.322$ & $0.081$  \\
\textrm{combined}        & $0.846$ & $0.030$ & $0.308$ & $0.032$  \\
\noalign{\vspace{0.2cm}}
\hline

\toprule
\rowcolor{Gray}
\multicolumn{5}{c}{$M_\textrm{min} = 5\cdot10^{12} {M_\odot /}{h}$}  \\
\hline 
\noalign{\smallskip}
{} & \multicolumn{2}{c}{$\sigma_8$} & \multicolumn{2}{c}{$\Omega_{\rm M}$}\\
\noalign{\smallskip}
\cline{2-3}\cline{4-5}
{} & \textrm{Mean} & \textrm{St. dev.} & \textrm{Mean} & \textrm{St. dev.} \\
\cline{2-5}
\noalign{\smallskip}
$z=0.00$                 & $0.853$ & $0.051$ & $0.295$ & $0.059$  \\
$z=0.55$                 & $0.901$ & $0.124$ & $0.333$ & $0.093$  \\
$z=1.00$                 & $0.989$ & $0.204$ & $0.360$ & $0.116$  \\
\textrm{combined}        & $0.856$ & $0.045$ & $0.293$ & $0.041$  \\
\noalign{\vspace{0.2cm}}
\hline

\toprule
\rowcolor{Gray}
\multicolumn{5}{c}{$M_\textrm{min} = 1\cdot10^{13} {M_\odot /}{h}$}  \\
\hline 
\noalign{\smallskip}
{} & \multicolumn{2}{c}{$\sigma_8$} & \multicolumn{2}{c}{$\Omega_{\rm M}$}\\
\noalign{\smallskip}
\cline{2-3}\cline{4-5}
{} & \textrm{Mean} & \textrm{St. dev.} & \textrm{Mean} & \textrm{St. dev.} \\
\cline{2-5}
\noalign{\smallskip}
$z=0.00$                 & $0.802$ & $0.062$ & $0.228$ & $0.054$  \\
$z=0.55$                 & $0.910$ & $0.147$ & $0.310$ & $0.091$  \\
$z=1.00$                 & $1.047$ & $0.247$ & $0.378$ & $0.125$  \\
\textrm{combined}        & $0.822$ & $0.053$ & $0.256$ & $0.040$  \\
\noalign{\smallskip}
\hline
\bottomrule

\end{tabularx}
\label{tab:parameters_combined}
\end{table}

\begin{figure}
\plotone{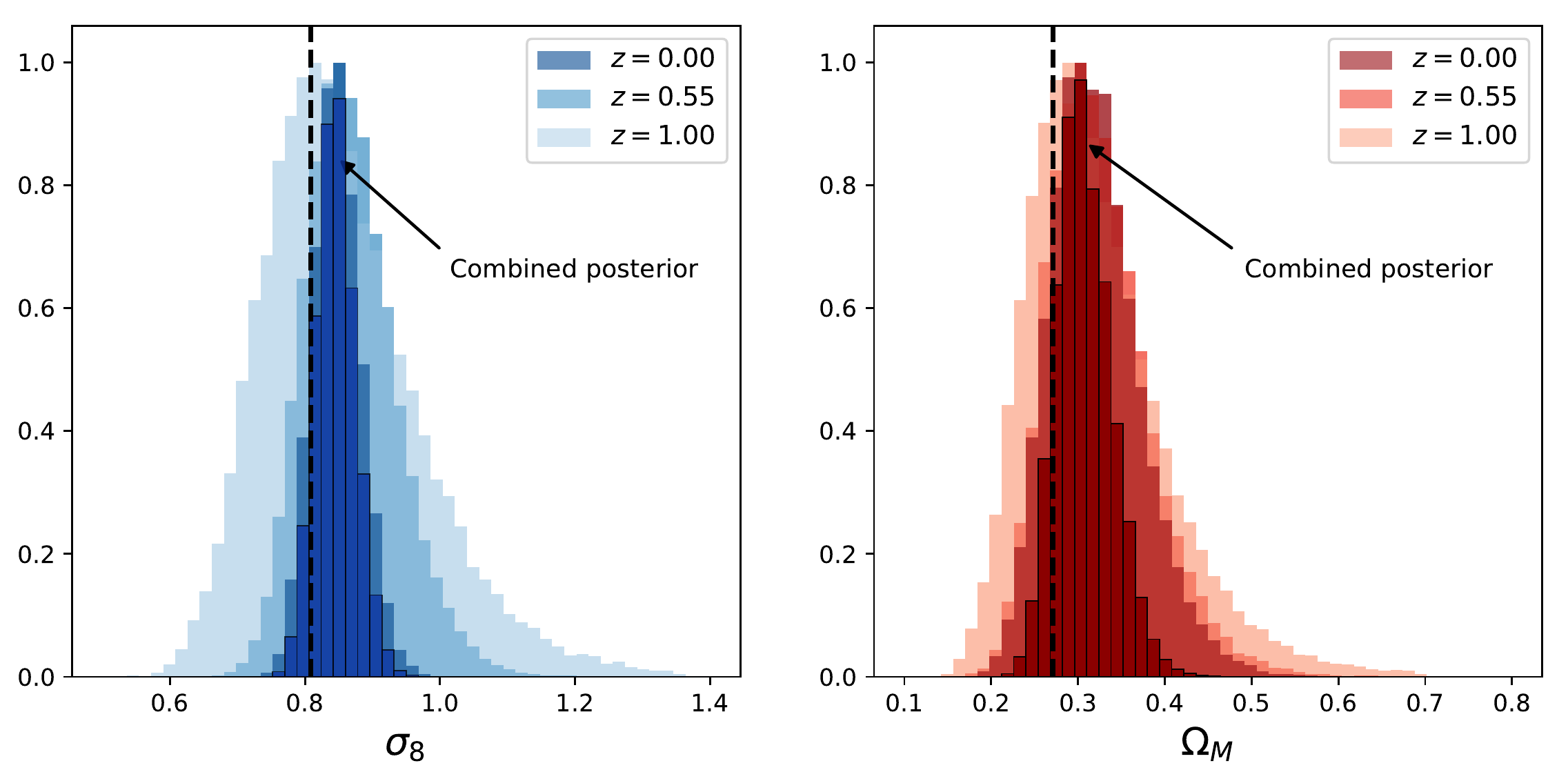}
    \caption{Normalised posterior probabilities of $\sigma_8$ (\textit{left}) and $\Omega_{\rm M}$ (\textit{right}) computed for the halo catalogue with $M_{\textrm{min}} = 2 \cdot 10^{12} \ {M_\odot /}{h}$ at redshift $z=0$, $z=0.55$ and $z=1$. The histograms with black outlines represent the combined distributions achieved by multiplying all the posterior probabilities relative to different redshifts. The black dashed lines indicate the true WMAP7 values ($\sigma_8=0.809$ and $\Omega_{\rm M}=0.2711$).}
    \label{histos}
\end{figure}

In this Appendix, we test the constraining power given by the combination of the posterior probabilities obtained performing the Bayesian statistical analysis of the measured void size functions for a DM halo catalogue at different redshifts.
Thanks to the redshift dependence of the degeneracy directions, it is possible to derive tighter constraints on the values achieved for $\sigma_8$ and $\Omega_{\rm M}$. Despite our samples cannot be considered completely independent, we multiply the posterior probabilities at different redshifts as if they were achieved from independent data, in order to reproduce the results that would be obtained from separate redshift shells in real surveys.
We show in Fig. \ref{histos} the results for the halo catalogue with $M_{\textrm{min}} = 2 \cdot 10^{12} \ {M_\odot /}{h}$ obtained by multiplying the posterior distributions for the parameters at $z=0$, $z=0.55$ and $z=1$. Table \ref{tab:parameters_combined} reports the mean values and the standard deviations of the posterior distributions of $\sigma_8$ and $\Omega_{\rm M}$ at these redshifts also for the catalogues with  $M_\textrm{min} = 5 \cdot 10^{12} \ {M_\odot /}{h}$ and $10^{13} \ {M_\odot /}{h}$, together with the analogous quantities obtained for the combined posterior probability. As expected, by joining the information at different redshifts, we can achieve more precise constraints on the cosmological parameters, as shown by the decreasing of the width of the combined posterior distributions.


\bibliographystyle{mn2e} 
\bibliography{bib}

\begin{thebibliography}{70}
\expandafter\ifx\csname natexlab\endcsname\relax\def\natexlab#1{#1}\fi

\bibitem[{{Achitouv}(2017)}]{Achitouv2017}
{Achitouv} I., 2017, \prd, 96, 083506

\bibitem[{{Achitouv}, {Neyrinck} \& {Paranjape}(2015){Achitouv}, {Neyrinck}, \&
  {Paranjape}}]{Achitouv2015}
{Achitouv} I., {Neyrinck} M., {Paranjape} A., 2015, \mnras, 451, 3964

\bibitem[{{Adermann} {et~al}\mbox{.}(2017){Adermann}, {Elahi}, {Lewis}, \&
  {Power}}]{Adermann2017}
{Adermann} E., {Elahi} P.~J., {Lewis} G.~F., {Power} C., 2017, \mnras, 468,
  3381

\bibitem[{{Adermann} {et~al}\mbox{.}(2018){Adermann}, {Elahi}, {Lewis}, \&
  {Power}}]{Adermann2018}
{Adermann} E., {Elahi} P.~J., {Lewis} G.~F., {Power} C., 2018, \mnras, 479,
  4861

\bibitem[{{Alcock} \& {Paczynski}(1979)}]{AP1979}
{Alcock} C., {Paczynski} B., 1979, \nat, 281, 358

\bibitem[{{Baldi}(2012)}]{baldi2012}
{Baldi} M., 2012, \mnras, 422, 1028

\bibitem[{{Baldi} {et~al}\mbox{.}(2010){Baldi}, {Pettorino}, {Robbers}, \&
  {Springel}}]{baldi2010}
{Baldi} M., {Pettorino} V., {Robbers} G., {Springel} V., 2010, \mnras, 403,
  1684

\bibitem[{{Barreira} {et~al}\mbox{.}(2015){Barreira}, {Cautun}, {Li}, {Baugh},
  \& {Pascoli}}]{Barreira2015}
{Barreira} A., {Cautun} M., {Li} B., {Baugh} C.~M., {Pascoli} S., 2015, \jcap,
  8, 028

\bibitem[{{Bernardeau}(1994)}]{bernardeau}
{Bernardeau} F., 1994, \apj, 427, 51

\bibitem[{{Bond} {et~al}\mbox{.}(1991){Bond}, {Cole}, {Efstathiou}, \&
  {Kaiser}}]{Bond}
{Bond} J.~R., {Cole} S., {Efstathiou} G., {Kaiser} N., 1991, \apj, 379, 440

\bibitem[{{Bos} {et~al}\mbox{.}(2012){Bos}, {van de Weygaert}, {Dolag}, \&
  {Pettorino}}]{Bos2012}
{Bos} E.~G.~P., {van de Weygaert} R., {Dolag} K., {Pettorino} V., 2012, \mnras,
  426, 440

\bibitem[{{Cai}, {Padilla} \& {Li}(2015){Cai}, {Padilla}, \& {Li}}]{cai2015}
{Cai} Y.-C., {Padilla} N., {Li} B., 2015, \mnras, 451, 1036

\bibitem[{{Cai} {et~al}\mbox{.}(2016){Cai}, {Taylor}, {Peacock}, \&
  {Padilla}}]{cai2016}
{Cai} Y.-C., {Taylor} A., {Peacock} J.~A., {Padilla} N., 2016, \mnras, 462,
  2465

\bibitem[{{Clampitt}, {Cai} \& {Li}(2013){Clampitt}, {Cai}, \&
  {Li}}]{Clampitt2013}
{Clampitt} J., {Cai} Y.-C., {Li} B., 2013, \mnras, 431, 749

\bibitem[{{Clampitt} \& {Jain}(2015)}]{clampitt2015}
{Clampitt} J., {Jain} B., 2015, \mnras, 454, 3357

\bibitem[{{Colberg} {et~al}\mbox{.}(2008){Colberg}, {Pearce}, {Foster},
  {Platen}, {Brunino}, {Neyrinck}, {Basilakos}, {Fairall}, {Feldman},
  {Gottl{\"o}ber}, {Hahn}, {Hoyle}, {M{\"u}ller}, {Nelson}, {Plionis},
  {Porciani}, {Shandarin}, {Vogeley}, \& {van de Weygaert}}]{Colberg_finders}
{Colberg} J.~M. {et~al.}, 2008, \mnras, 387, 933

\bibitem[{{Davies}, {Cautun} \& {Li}(2018){Davies}, {Cautun}, \&
  {Li}}]{Davies2018}
{Davies} C.~T., {Cautun} M., {Li} B., 2018, \mnras, 480, L101

\bibitem[{{Dekel} \& {Lahav}(1999)}]{Dekel1999}
{Dekel} A., {Lahav} O., 1999, \apj, 520, 24

\bibitem[{{Di Porto} {et~al}\mbox{.}(2016){Di Porto}, {Branchini}, {Bel},
  {Marulli}, {Bolzonella}, {Cucciati}, {de la Torre}, {Granett}, {Guzzo},
  {Marinoni}, {Moscardini}, {Abbas}, {Adami}, {Arnouts}, {Bottini}, {Cappi},
  {Coupon}, {Davidzon}, {De Lucia}, {Fritz}, {Franzetti}, {Fumana}, {Garilli},
  {Ilbert}, {Iovino}, {Krywult}, {Le Brun}, {Le F{\`e}vre}, {Maccagni},
  {Ma{\l}ek}, {McCracken}, {Paioro}, {Polletta}, {Pollo}, {Scodeggio}, {Tasca},
  {Tojeiro}, {Vergani}, {Zanichelli}, {Burden}, {Marchetti}, {Martizzi},
  {Mellier}, {Nichol}, {Peacock}, {Percival}, {Viel}, {Wolk}, \&
  {Zamorani}}]{DiPorto2016}
{Di Porto} C. {et~al.}, 2016, \aap, 594, A62

\bibitem[{{Elyiv} {et~al}\mbox{.}(2015){Elyiv}, {Marulli}, {Pollina}, {Baldi},
  {Branchini}, {Cimatti}, \& {Moscardini}}]{Elyiv2015}
{Elyiv} A., {Marulli} F., {Pollina} G., {Baldi} M., {Branchini} E., {Cimatti}
  A., {Moscardini} L., 2015, \mnras, 448, 642

\bibitem[{{Falck} {et~al}\mbox{.}(2018){Falck}, {Koyama}, {Zhao}, \&
  {Cautun}}]{Falck2018}
{Falck} B., {Koyama} K., {Zhao} G.-B., {Cautun} M., 2018, \mnras, 475, 3262

\bibitem[{{Hamaus} {et~al}\mbox{.}(2017){Hamaus}, {Cousinou}, {Pisani},
  {Aubert}, {Escoffier}, \& {Weller}}]{Hamaus2017}
{Hamaus} N., {Cousinou} M.-C., {Pisani} A., {Aubert} M., {Escoffier} S.,
  {Weller} J., 2017, \jcap, 7, 014

\bibitem[{{Hamaus} {et~al}\mbox{.}(2016){Hamaus}, {Pisani}, {Sutter}, {Lavaux},
  {Escoffier}, {Wandelt}, \& {Weller}}]{Hamaus2016}
{Hamaus} N., {Pisani} A., {Sutter} P.~M., {Lavaux} G., {Escoffier} S.,
  {Wandelt} B.~D., {Weller} J., 2016, Physical Review Letters, 117, 091302

\bibitem[{{Hamaus} {et~al}\mbox{.}(2014){Hamaus}, {Sutter}, {Lavaux}, \&
  {Wandelt}}]{hamaus2014}
{Hamaus} N., {Sutter} P.~M., {Lavaux} G., {Wandelt} B.~D., 2014, \jcap, 12, 013

\bibitem[{{Hamaus} {et~al}\mbox{.}(2015){Hamaus}, {Sutter}, {Lavaux}, \&
  {Wandelt}}]{Hamaus2015}
{Hamaus} N., {Sutter} P.~M., {Lavaux} G., {Wandelt} B.~D., 2015, \jcap, 11, 036

\bibitem[{{Hawken} {et~al}\mbox{.}(2017){Hawken}, {Granett}, {Iovino}, {Guzzo},
  {Peacock}, {de la Torre}, {Garilli}, {Bolzonella}, {Scodeggio}, {Abbas},
  {Adami}, {Bottini}, {Cappi}, {Cucciati}, {Davidzon}, {Fritz}, {Franzetti},
  {Krywult}, {Le Brun}, {Le F{\`e}vre}, {Maccagni}, {Ma{\l}ek}, {Marulli},
  {Polletta}, {Pollo}, {Tasca}, {Tojeiro}, {Vergani}, {Zanichelli}, {Arnouts},
  {Bel}, {Branchini}, {De Lucia}, {Ilbert}, {Moscardini}, \&
  {Percival}}]{Hawken}
{Hawken} A.~J. {et~al.}, 2017, \aap, 607, A54

\bibitem[{{Icke}(1984)}]{icke1984voids}
{Icke} V., 1984, \mnras, 206, 1P

\bibitem[{{Jennings}, {Li} \& {Hu}(2013){Jennings}, {Li}, \& {Hu}}]{Jennings}
{Jennings} E., {Li} Y., {Hu} W., 2013, \mnras, 434, 2167

\bibitem[{{Komatsu} {et~al}\mbox{.}(2011){Komatsu} {et~al.}}]{komatsu2011}
{Komatsu} E., {et~al.}, 2011, \apjs, 192, 18

\bibitem[{{Kreisch} {et~al}\mbox{.}(2019){Kreisch}, {Pisani}, {Carbone}, {Liu},
  {Hawken}, {Massara}, {Spergel}, \& {Wandelt}}]{Kreisch2018}
{Kreisch} C.~D., {Pisani} A., {Carbone} C., {Liu} J., {Hawken} A.~J., {Massara}
  E., {Spergel} D.~N., {Wandelt} B.~D., 2019, \mnras, 1877

\bibitem[{{Landy} \& {Szalay}(1993)}]{Landy_Szalay1993}
{Landy} S.~D., {Szalay} A.~S., 1993, \apj, 412, 64

\bibitem[{{Lavaux} \& {Wandelt}(2012)}]{Lav&Wan2012}
{Lavaux} G., {Wandelt} B.~D., 2012, \apj, 754, 109

\bibitem[{{Lee} \& {Park}(2009)}]{2009Lee}
{Lee} J., {Park} D., 2009, \apj, 696, L10

\bibitem[{{Maartens}, {Clarkson} \& {Chen}(2018){Maartens}, {Clarkson}, \&
  {Chen}}]{negative_bias2}
{Maartens} R., {Clarkson} C., {Chen} S., 2018, \jcap, 2018, 013

\bibitem[{{Marulli} {et~al}\mbox{.}(2013){Marulli}, {Bolzonella}, {Branchini},
  {Davidzon}, {de la Torre}, {Granett}, {Guzzo}, {Iovino}, {Moscardini},
  {Pollo}, {Abbas}, {Adami}, {Arnouts}, {Bel}, {Bottini}, {Cappi}, {Coupon},
  {Cucciati}, {De Lucia}, {Fritz}, {Franzetti}, {Fumana}, {Garilli}, {Ilbert},
  {Krywult}, {Le Brun}, {Le F{\`e}vre}, {Maccagni}, {Ma{\l}ek}, {McCracken},
  {Paioro}, {Polletta}, {Schlagenhaufer}, {Scodeggio}, {Tasca}, {Tojeiro},
  {Vergani}, {Zanichelli}, {Burden}, {Di Porto}, {Marchetti}, {Marinoni},
  {Mellier}, {Nichol}, {Peacock}, {Percival}, {Phleps}, {Wolk}, \&
  {Zamorani}}]{Marulli2013}
{Marulli} F. {et~al.}, 2013, \aap, 557, A17

\bibitem[{{Marulli}, {Veropalumbo} \& {Moresco}(2016){Marulli}, {Veropalumbo},
  \& {Moresco}}]{CBL}
{Marulli} F., {Veropalumbo} A., {Moresco} M., 2016, Astronomy and Computing,
  14, 35

\bibitem[{{Marulli} {et~al}\mbox{.}(2018){Marulli}, {Veropalumbo}, {Sereno},
  {Moscardini}, {Pacaud}, {Pierre}, {Plionis}, {Cappi}, {Adami}, {Alis},
  {Altieri}, {Birkinshaw}, {Ettori}, {Faccioli}, {Gastaldello}, {Koulouridis},
  {Lidman}, {Le F{\`e}vre}, {Maurogordato}, {Poggianti}, {Pompei},
  {Sadibekova}, \& {Valtchanov}}]{Marulli2018}
{Marulli} F. {et~al.}, 2018, \aap, 620, A1

\bibitem[{{Massara} {et~al}\mbox{.}(2015){Massara}, {Villaescusa-Navarro},
  {Viel}, \& {Sutter}}]{massara2015}
{Massara} E., {Villaescusa-Navarro} F., {Viel} M., {Sutter} P.~M., 2015, \jcap,
  11, 018

\bibitem[{{Melchior} {et~al}\mbox{.}(2014){Melchior}, {Sutter}, {Sheldon},
  {Krause}, \& {Wandelt}}]{Melchior2014}
{Melchior} P., {Sutter} P.~M., {Sheldon} E.~S., {Krause} E., {Wandelt} B.~D.,
  2014, \mnras, 440, 2922

\bibitem[{{Micheletti} {et~al}\mbox{.}(2014){Micheletti}, {Iovino}, {Hawken},
  {Granett}, {Bolzonella}, {Cappi}, {Guzzo}, {Abbas}, {Adami}, {Arnouts},
  {Bel}, {Bottini}, {Branchini}, {Coupon}, {Cucciati}, {Davidzon}, {De Lucia},
  {de la Torre}, {Fritz}, {Franzetti}, {Fumana}, {Garilli}, {Ilbert},
  {Krywult}, {Le Brun}, {Le F{\`e}vre}, {Maccagni}, {Ma{\l}ek}, {Marulli},
  {McCracken}, {Polletta}, {Pollo}, {Schimd}, {Scodeggio}, {Tasca}, {Tojeiro},
  {Vergani}, {Zanichelli}, {Burden}, {Di Porto}, {Marchetti}, {Marinoni},
  {Mellier}, {Moutard}, {Moscardini}, {Nichol}, {Peacock}, {Percival}, \&
  {Zamorani}}]{micheletti_vimos}
{Micheletti} D. {et~al.}, 2014, \aap, 570, A106

\bibitem[{{Nadathur}, {Carter} \& {Percival}(2019){Nadathur}, {Carter}, \&
  {Percival}}]{Nadathur2019a}
{Nadathur} S., {Carter} P., {Percival} W.~J., 2019, \mnras, 482, 2459

\bibitem[{{Nadathur} \& {Hotchkiss}(2015{\natexlab{a}})}]{NadathurI}
{Nadathur} S., {Hotchkiss} S., 2015{\natexlab{a}}, \mnras, 454, 2228

\bibitem[{{Nadathur} \& {Hotchkiss}(2015{\natexlab{b}})}]{NadathurII}
{Nadathur} S., {Hotchkiss} S., 2015{\natexlab{b}}, \mnras, 454, 889

\bibitem[{{Nadathur} \& {Percival}(2019)}]{Natathur2019b}
{Nadathur} S., {Percival} W.~J., 2019, \mnras, 483, 3472

\bibitem[{{Neyrinck}(2008)}]{2008MNRAS.386.2101N}
{Neyrinck} M.~C., 2008, \mnras, 386, 2101

\bibitem[{{Norberg} {et~al}\mbox{.}(2009){Norberg}, {Baugh}, {Gazta{\~n}aga},
  \& {Croton}}]{Norberg2019}
{Norberg} P., {Baugh} C.~M., {Gazta{\~n}aga} E., {Croton} D.~J., 2009, \mnras,
  396, 19

\bibitem[{{P{\'e}nin}, {Umeh} \& {Santos}(2018){P{\'e}nin}, {Umeh}, \&
  {Santos}}]{negative_bias1}
{P{\'e}nin} A., {Umeh} O., {Santos} M.~G., 2018, \mnras, 473, 4297

\bibitem[{{Pisani} {et~al}\mbox{.}(2015){Pisani}, {Sutter}, {Hamaus},
  {Alizadeh}, {Biswas}, {Wandelt}, \& {Hirata}}]{Pisani}
{Pisani} A., {Sutter} P.~M., {Hamaus} N., {Alizadeh} E., {Biswas} R., {Wandelt}
  B.~D., {Hirata} C.~M., 2015, \prd, 92, 083531

\bibitem[{{Platen}, {van de Weygaert} \& {Jones}(2007){Platen}, {van de
  Weygaert}, \& {Jones}}]{2007MNRAS.380..551P}
{Platen} E., {van de Weygaert} R., {Jones} B.~J.~T., 2007, \mnras, 380, 551

\bibitem[{{Pollina} {et~al}\mbox{.}(2016){Pollina}, {Baldi}, {Marulli}, \&
  {Moscardini}}]{Pollina2016}
{Pollina} G., {Baldi} M., {Marulli} F., {Moscardini} L., 2016, \mnras, 455,
  3075

\bibitem[{{Pollina} {et~al}\mbox{.}(2017){Pollina}, {Hamaus}, {Dolag},
  {Weller}, {Baldi}, \& {Moscardini}}]{pollinalinear}
{Pollina} G., {Hamaus} N., {Dolag} K., {Weller} J., {Baldi} M., {Moscardini}
  L., 2017, \mnras, 469, 787

\bibitem[{{Pollina} {et~al}\mbox{.}(2018){Pollina}, {Hamaus}, {Paech}, {Dolag},
  {Weller}, {S{\'a}nchez}, {Rykoff}, {Jain}, {Abbott}, {Allam}, {Avila},
  {Bernstein}, {Bertin}, {Brooks}, {Burke}, {Carnero Rosell}, {Carrasco Kind},
  {Carretero}, {Cunha}, {D'Andrea}, {da Costa}, {De Vicente}, {DePoy}, {Desai},
  {Diehl}, {Doel}, {Evrard}, {Flaugher}, {Fosalba}, {Frieman},
  {Garc{\'{\i}}a-Bellido}, {Gerdes}, {Giannantonio}, {Gruen}, {Gschwend},
  {Gutierrez}, {Hartley}, {Hollowood}, {Honscheid}, {Hoyle}, {James},
  {Jeltema}, {Kuehn}, {Kuropatkin}, {Lima}, {March}, {Marshall}, {Melchior},
  {Menanteau}, {Miquel}, {Plazas}, {Romer}, {Sanchez}, {Scarpine}, {Schindler},
  {Schubnell}, {Sevilla-Noarbe}, {Smith}, {Soares-Santos}, {Sobreira},
  {Suchyta}, {Tarle}, {Walker}, \& {Wester}}]{pollinarelative}
{Pollina} G. {et~al.}, 2018, ArXiv e-prints: 1806.06860

\bibitem[{{Press} \& {Schechter}(1974)}]{P&S}
{Press} W.~H., {Schechter} P., 1974, \apj, 193, 437

\bibitem[{{Pycke} \& {Russell}(2016)}]{pycke&russel2016}
{Pycke} J.-R., {Russell} E., 2016, \apj, 821, 110

\bibitem[{{Ronconi} {et~al}\mbox{.}(2019){Ronconi}, {Contarini}, {Marulli},
  {Baldi}, \& {Moscardini}}]{Roncarini2019}
{Ronconi} T., {Contarini} S., {Marulli} F., {Baldi} M., {Moscardini} L., 2019,
  arXiv e-prints: 1902.04585

\bibitem[{{Ronconi} \& {Marulli}(2017)}]{Ronconi2017}
{Ronconi} T., {Marulli} F., 2017, \aap, 607, A24

\bibitem[{{Sahl{\'e}n}(2019)}]{martinDE&neutrinos_prop}
{Sahl{\'e}n} M., 2019, \prd, 99, 063525

\bibitem[{{Sahl{\'e}n} \& {Silk}(2018)}]{Martin_modifiedG}
{Sahl{\'e}n} M., {Silk} J., 2018, \prd, 97, 103504

\bibitem[{{S{\'a}nchez} {et~al}\mbox{.}(2017){S{\'a}nchez}, {Clampitt},
  {Kovacs}, {Jain}, {Garc{\'{\i}}a-Bellido}, {Nadathur}, {Gruen}, {Hamaus},
  {Huterer}, {Vielzeuf}, {Amara}, {Bonnett}, {DeRose}, {Hartley}, {Jarvis},
  {Lahav}, {Miquel}, {Rozo}, {Rykoff}, {Sheldon}, {Wechsler}, {Zuntz},
  {Abbott}, {Abdalla}, {Annis}, {Benoit-L{\'e}vy}, {Bernstein}, {Bernstein},
  {Bertin}, {Brooks}, {Buckley-Geer}, {Carnero Rosell}, {Carrasco Kind},
  {Carretero}, {Crocce}, {Cunha}, {D'Andrea}, {da Costa}, {Desai}, {Diehl},
  {Dietrich}, {Doel}, {Evrard}, {Fausti Neto}, {Flaugher}, {Fosalba},
  {Frieman}, {Gaztanaga}, {Gruendl}, {Gutierrez}, {Honscheid}, {James},
  {Krause}, {Kuehn}, {Lima}, {Maia}, {Marshall}, {Melchior}, {Plazas}, {Reil},
  {Romer}, {Sanchez}, {Schubnell}, {Sevilla-Noarbe}, {Smith}, {Soares-Santos},
  {Sobreira}, {Suchyta}, {Tarle}, {Thomas}, {Walker}, {Weller}, \& {DES
  Collaboration}}]{sanchez2017}
{S{\'a}nchez} C. {et~al.}, 2017, \mnras, 465, 746

\bibitem[{{Sheth} \& {van de Weygaert}(2004)}]{Sheth}
{Sheth} R.~K., {van de Weygaert} R., 2004, \mnras, 350, 517

\bibitem[{{Sutter} {et~al}\mbox{.}(2015){Sutter}, {Lavaux}, {Hamaus}, {Pisani},
  {Wandelt}, {Warren}, {Villaescusa-Navarro}, {Zivick}, {Mao}, \&
  {Thompson}}]{SutterVIDE}
{Sutter} P.~M. {et~al.}, 2015, Astronomy and Computing, 9, 1

\bibitem[{{Sutter} {et~al}\mbox{.}(2014){Sutter}, {Lavaux}, {Hamaus},
  {Wandelt}, {Weinberg}, \& {Warren}}]{sutter2014}
{Sutter} P.~M., {Lavaux} G., {Hamaus} N., {Wandelt} B.~D., {Weinberg} D.~H.,
  {Warren} M.~S., 2014, \mnras, 442, 462

\bibitem[{{Sutter} {et~al}\mbox{.}(2012){Sutter}, {Lavaux}, {Wandelt}, \&
  {Weinberg}}]{sutter2012}
{Sutter} P.~M., {Lavaux} G., {Wandelt} B.~D., {Weinberg} D.~H., 2012, \apj,
  761, 187

\bibitem[{{Szapudi} {et~al}\mbox{.}(2015){Szapudi}, {Kov{\'a}cs}, {Granett},
  {Frei}, {Silk}, {Burgett}, {Cole}, {Draper}, {Farrow}, {Kaiser}, {Magnier},
  {Metcalfe}, {Morgan}, {Price}, {Tonry}, \& {Wainscoat}}]{Szapudi}
{Szapudi} I. {et~al.}, 2015, \mnras, 450, 288

\bibitem[{{Tikhonov} \& {Karachentsev}(2006)}]{Tikhonov}
{Tikhonov} A.~V., {Karachentsev} I.~D., 2006, \apj, 653, 969

\bibitem[{{Tinker} {et~al}\mbox{.}(2010){Tinker}, {Robertson}, {Kravtsov},
  {Klypin}, {Warren}, {Yepes}, \& {Gottl{\"o}ber}}]{Tinker_bias}
{Tinker} J.~L., {Robertson} B.~E., {Kravtsov} A.~V., {Klypin} A., {Warren}
  M.~S., {Yepes} G., {Gottl{\"o}ber} S., 2010, \apj, 724, 878

\bibitem[{{van de Weygaert} \& {van Kampen}(1993)}]{vdW&kampen}
{van de Weygaert} R., {van Kampen} E., 1993, \mnras, 263, 481

\bibitem[{{Verza} {et~al}\mbox{.}(2019){Verza}, {Pisani}, {Carbone}, {Hamaus},
  \& {Guzzo}}]{verza2019}
{Verza} G., {Pisani} A., {Carbone} C., {Hamaus} N., {Guzzo} L., 2019, arXiv
  e-prints, arXiv:1906.00409

\bibitem[{{Wojtak}, {Powell} \& {Abel}(2016){Wojtak}, {Powell}, \&
  {Abel}}]{Wojtak2016}
{Wojtak} R., {Powell} D., {Abel} T., 2016, \mnras, 458, 4431

\bibitem[{{Zivick} {et~al}\mbox{.}(2015){Zivick}, {Sutter}, {Wandelt}, {Li}, \&
  {Lam}}]{Zivick2015}
{Zivick} P., {Sutter} P.~M., {Wandelt} B.~D., {Li} B., {Lam} T.~Y., 2015,
  \mnras, 451, 4215

\end{thebibliography}

\end{document}